\documentclass[12pt,preprint]{aastex}
\usepackage{graphicx,color}
\newcommand{\cophc}{{CoPhC}} 
\newcommand{\jcoph}{{JCoPh}} 
\newcommand{\jgra}{{JGRA}} 
\newcommand{\natur}{{Natur}} 
\newcommand{\phpl}{{PhPl}} 
\newcommand{\sci}{{Sci}} 
\newcommand{\soph}{{SoPh}} 
\newcommand{\ssrv}{{SSRv}} 
\shorttitle{Escape of Flare-Accelerated Particles in Solar Eruptive Events}
\shortauthors{Masson, Antiochos, \& DeVore}

\begin{document}

\title{Escape of Flare-accelerated Particles in Solar Eruptive Events}

\author{S.\ Masson}
 \affil{LESIA, Observatoire de Paris, PSL Research University, CNRS, Sorbonne Universités, UPMC Univ.\ Paris 06, Univ.\ Paris Diderot,
Sorbonne Paris Cité, France}

\email{sophie.masson@obspm.fr} 


\author{S.\ K.\ Antiochos and C.\ R.\ DeVore}
 \affil{Heliophysics Science Division, NASA Goddard Space Flight Center, 8800 Greenbelt Road, Greenbelt MD 20771 USA} 

 

\begin{abstract}

Impulsive solar energetic particle events are widely believed to be due to the prompt escape into the interplanetary medium of flare-accelerated particles produced by solar eruptive events. 
According to the standard model for such events, however, particles accelerated by the flare reconnection should remain trapped in the flux rope comprising the coronal mass ejection. The particles should reach the Earth only much later, along with the bulk ejecta. To resolve this paradox, we have extended our previous axisymmetric model for the escape of flare-accelerated particles to fully three-dimensional (3D) geometries. We report the results of magnetohydrodynamic simulations of a coronal system that consists of a bipolar active region embedded in a background global dipole field structured by solar wind. Our simulations show that multiple magnetic reconnection episodes occur prior to and during the CME eruption and its interplanetary propagation. In addition to the episodes that build up the flux rope, reconnection between the open field and the CME couples the closed corona to the open interplanetary field. Flare-accelerated particles initially trapped in the CME thereby gain access to the open interplanetary field along a trail blazed by magnetic reconnection. A key difference between these 3D results and our previous calculations is that the interchange reconnection allows accelerated particles to escape from deep within the CME flux-rope. We estimate the spatial extent of the particle-escape channels. The relative timings between flare acceleration and release of the energetic particles through CME/open-field coupling are also determined. All our results compare favourably with observations.

\end{abstract}



\section{Introduction}
\label{intro} 

A significant fraction of the energy released during solar eruptions is 
transferred to particles that are accelerated to high energy \citep{Cargill_al12}. 
Some of those particles precipitate towards the solar surface, producing
extreme ultraviolet (EUV), X-ray, and even $\gamma$-ray emissions,
while others escape into the interplanetary medium to be detected by
in situ instruments. The mechanisms responsible for the acceleration
of those particles, in particular the most energetic ones with
relativistic energies, are far from fully understood.

Candidate mechanisms for explaining the acceleration of solar
energetic particles (SEPs) include the shock wave driven by the
coronal mass ejection (CME) and small-scale processes that operate
during magnetic reconnection associated with the flare. SEP events 
are of two types. Gradual events are thought to be produced by 
energetic particles accelerated at the CME-driven shock \citep{Reames99}. 
Impulsive events are regarded as being due to particles accelerated 
by magnetic reconnection during the flare \citep[hereafter 
called flare-accelerated particles;][]{Cane_al86}. Historically,
the major difference between the gradual and impulsive events is that
the graduals are associated with eruptive flares occurring deep in the
magnetically closed corona, while the impulsives are associated with
flares near open-field regions observed as coronal holes \citep{Pick_al06}.

In order to reach the interplanetary medium and the Earth, the
energetic particles must be transported from their acceleration site
to interplanetary magnetic field (IMF) lines that connect to
Earth and along which the particles propagate to Earth. For energetic
particles accelerated by the shock wave at the front of a CME, the
injection occurs when the particles reach a sufficiently high velocity
to escape from the shocking region. Therefore, these escaped particles
are injected from the shocked regions onto the IMF field lines just
ahead of and behind the CME shock. The particle injection from the
shock to the interplanetary medium has been confirmed by a
multi-instrument analysis of a particular SEP event detected at
multiple longitudinal positions
\citep{Rouillard_al11,Richardson_al14,Lario_al14}.

Gradual events are usually attributed to shock-accelerated particles
\citep{Rouillard_al11,Richardson_al14,Lario_al14}. However, to
determine the SEP acceleration mechanisms, one uses observational
diagnostics combining in situ measurements of energetic particle
fluxes with remote-sensing observations of solar activity. When doing
so, the injection and propagation effects between the acceleration
region and the spacecraft are mainly ignored, introducing a bias in
the observational diagnostics. Some detailed studies of
relativistic-particle events have shown that energetic particles
detected at Earth during gradual events have been flare-accelerated
\citep{Klein_al99,Miroshnichenko_al05,
  Masson_al09a,Klein_al14,Lario_al17}. In those cases, the
flare-accelerated particles become SEPs at 1 AU, and somehow have
managed to access the open IMF. Particles accelerated at the 
interchange reconnection site between closed and open structures 
gain direct access to the open interplanetary field \citep{Masson_al12a}. 
In contrast, flare-accelerated particles produced in a large-scale 
solar eruption do not have such direct access to the open
field. The standard 3D CSHKP model \citep{Carmichael64,
  Sturrock66,Hirayama74,KoppPneuman76} for eruptive flares describes
the global evolution of solar eruptions consisting of a flare and a
CME.  As described by \citet{Masson_al13} (see their Figure 1), 
particles accelerated at the flare site should remain trapped within 
the post-flare loops and the ejected flux rope of the CME. Therefore, 
the particles are unable to escape promptly into the interplanetary 
medium \citep[see review by][]{Reames13}.

Previous numerical simulations of the 3D CSHKP model have
  addressed many aspects of CME properties, such as the triggering
  mechanism \citep{TorokKliem07,Lynch_al08,Zuccarello_al15} and the
  flux-rope formation
  \citep[e.g.][]{Lynch_al09,Aulanier_al10,Leake_al14,Dahlin_al19}. However,
  those simulations were all performed in a closed, static corona.
  Three-dimensional numerical simulations of the initiation and
  propagation of CMEs into the solar wind that have been performed
  heretofore either: 1) assumed a pre-existing flux rope to generate a
  fast CME \citep[e.g.\ ][]{Cohen_al10,Lugaz_al11,Torok_al18}; or 2)
  resulted in a slow streamer-blowout CME that travels only slightly
  faster than the ambient solar wind \citep[e.g.\
  ][]{vanderHolst_al09,Lynch_al16,Hosteaux_al18}. While reconnection
  undoubtedly occurred between the CME flux and the interplanetary
  field in the latter studies, the role of the reconnection in
  governing particle transport was not addressed. In contrast, in
  \citet{Masson_al13} and in this study, we specifically address the
  escape of flare-accelerated particles.  Our numerical simulations
  self-consistently produce a fast CME in response to subsonic
  footpoint motions applied to a potential magnetic configuration. The
  CME then propagates rapidly into, and interacts strongly with, the
  solar wind. We analyze, in detail, the reconnections between the
  escaping CME flux rope and surrounding open flux.

In \cite{Masson_al13}, hereafter MAD13, we developed an
axisymmetric (so-called 2.5D) model for the escape of flare-accelerated 
particles. It explains how particles energized at the flare site can 
be transported onto the open IMF, rather than remaining trapped in the 
closed field (whether the post-flare loops or the CME flux rope). 
Our model is based on the occurrence of so-called interchange
reconnection between the closed CME flux rope and the open IMF. This
reconnection allows the particles trapped within the CME to access the
IMF. The 2.5D geometry, however, is highly restrictive. Nonthermal particles
are observed to occur predominantly during the flare impulsive
phase, in which case the impulsive SEPs should be trapped deep in the
CME flux rope. We found that, in order for these SEPs to escape, the
rope was inevitably destroyed by interchange reconnection and did not
survive to become an interplanetary magnetic cloud, contrary to 
observations. Consequently, we have sought to extend the model to a 
full 3D geometry. Our goals are threefold. First, we 
confirm that the model remains valid in 3D. Second, we test 
whether 3D interchange reconnection can both release particles from 
the core of the CME and also preserve the integrity of the flux rope. 
Third, we determine the spatial and temporal characteristics 
of the escape of flare-accelerated particles for comparison with 
observations.

 
To address these objectives,
we performed and analyzed a large-scale 3D MHD simulation
of a non-symmetric magnetic configuration in which a bipolar active region is
embedded in the northern hemisphere of a solar background dipole
magnetic field.  The numerical model and the initial conditions are
described in \S\ref{model}. Then, we present our results on the
magnetic reconnection dynamics during the buildup to CME initiation
(\S\S\ref{null point reconnection},\ref{flux rope formation}), the
eruption itself (\S\ref{Eruption}), and the CME propagation into and
coupling with the solar wind (\S\ref{Coupling the CME and the
  interplanetary magnetic field}). We present the spatial and temporal
properties of the flare-accelerated particle escape in our 3D model in
\S\ref{Application to the escape of flare-accelerated particles}. We
discuss our results and their implications for particle escape and
future SEP studies in \S\ref{Discussion}. Finally, our summary
conclusions are given in \S\ref{Conclusion}.

\section{Model Description}
\label{model}

\subsection{Equations, Grids, and Boundary Conditions}
\label{arms}

The simulations were performed using the Adaptively Refined Magnetohydrodynamics Solver \citep[ARMS;][]{DeVoreAntiochos08}, solving the following ideal MHD equations in spherical coordinates:
\begin{eqnarray}
\label{eqcont}
\frac{\partial \rho}{\partial t} + {\bf \nabla} \cdot \left( \rho {\bf u} \right) &=&0,\\
\label{eqmom}
\frac{\partial \rho {\bf u}}{\partial t} +
  {\bf \nabla} \cdot \left( \rho{\bf u} {\bf u} \right) &=& (1/\mu) \left( {\bf \nabla} \times {\bf B} \right) \times {\bf B}  -  {\bf \nabla} P + \rho{\bf g} ,\\
\label{eqinduc}
 \frac{\partial {\bf B}}{\partial t} - {\bf \nabla} \times
  \left( {\bf u} \times {\bf B} \right) &=& 0,
\end{eqnarray}

\noindent where $\rho$ is the mass density, ${\bf u}$ the plasma
velocity, ${\bf B}$ the magnetic field, $P$ the pressure, $\mu$ the
permeability of free space, and ${\bf g} = - GM_\odot{\bf r}/r^3$ the
solar gravitational acceleration. We assume a fully ionized hydrogen
gas, so that the plasma pressure $P=2(\rho/m_p) k_B T$, where $T$ is
the temperature. For simplicity, we assume also that the temperature
is constant and uniform, $T \equiv T_0$. This isothermal model allows
high-lying field lines to be opened to the heliosphere by the solar
wind while permitting low-lying field lines to remain closed to the
Sun, as occurs ubiquitously in the corona. Our objective in this paper
is to simulate with high fidelity the changing connectivity of the
near-Sun magnetic field, not to predict the detailed thermodynamic
properties of the heliospheric plasma. Hence, we adopted the simplest
possible model that yields a solar wind and a dynamic,
self-consistently determined boundary between open and closed magnetic
structures in the corona.

The numerical scheme is a finite-volume multi-dimensional Flux
Corrected Transport algorithm \citep{DeVore91}. It ensures that the
divergence of the magnetic field remains small to machine accuracy,
and also that unphysical oscillations in all variables are minimized
while introducing minimal residual numerical diffusion. The equations
are solved using a second-order predictor-corrector in time and a
fourth-order integrator in space.  In conjunction with the flux
limiter, this accuracy is sufficient to inhibit numerical reconnection
until any developing current sheets (e.g., at the pre-existing null
point; \S\ref{topology}) thin down to the grid scale.

The spherical grid is stretched exponentially in radius $r$ and spaced uniformly in colatitude $\theta$ and longitude $\phi$. The computational domain covers the volume $r \in [1R_\odot,50R_\odot]$, $\theta \in [11.25^\circ,168.75^\circ]$, and $\phi \in [-180^\circ,+180^\circ]$. We use the same boundary conditions as in the 2.5D simulation, but extended to 3D \citep[for details, see][]{Masson_al13}. ARMS uses the parallel adaptive meshing toolkit PARAMESH \citep{MacNeice_al00} to tailor the grid to the evolving solution. In this simulation, the mesh was adaptively refined and coarsened over five levels of grids, determined by the ratio of the local scale of the three components of electric current density to the local grid spacing. Thus, the intense current sheets developing in the system are always resolved as finely as necessary on the mesh until the limiting resolution is reached. Although there is no explicit resistive term in the induction equation (\ref{eqinduc}), the numerical resistivity plays an important role in the intense, highly resolved current sheets. In addition to those dynamic current structures, we prescribed a maximum resolution layer of grid cells at the inner radial boundary within the dipolar active region, in order to resolve throughout the simulation the flows and gradients generated by the boundary forcing (\S\ref{forcing}).
 
\begin{figure}
\centerline{.pdf
\includegraphics[width=0.9\textwidth,clip=]{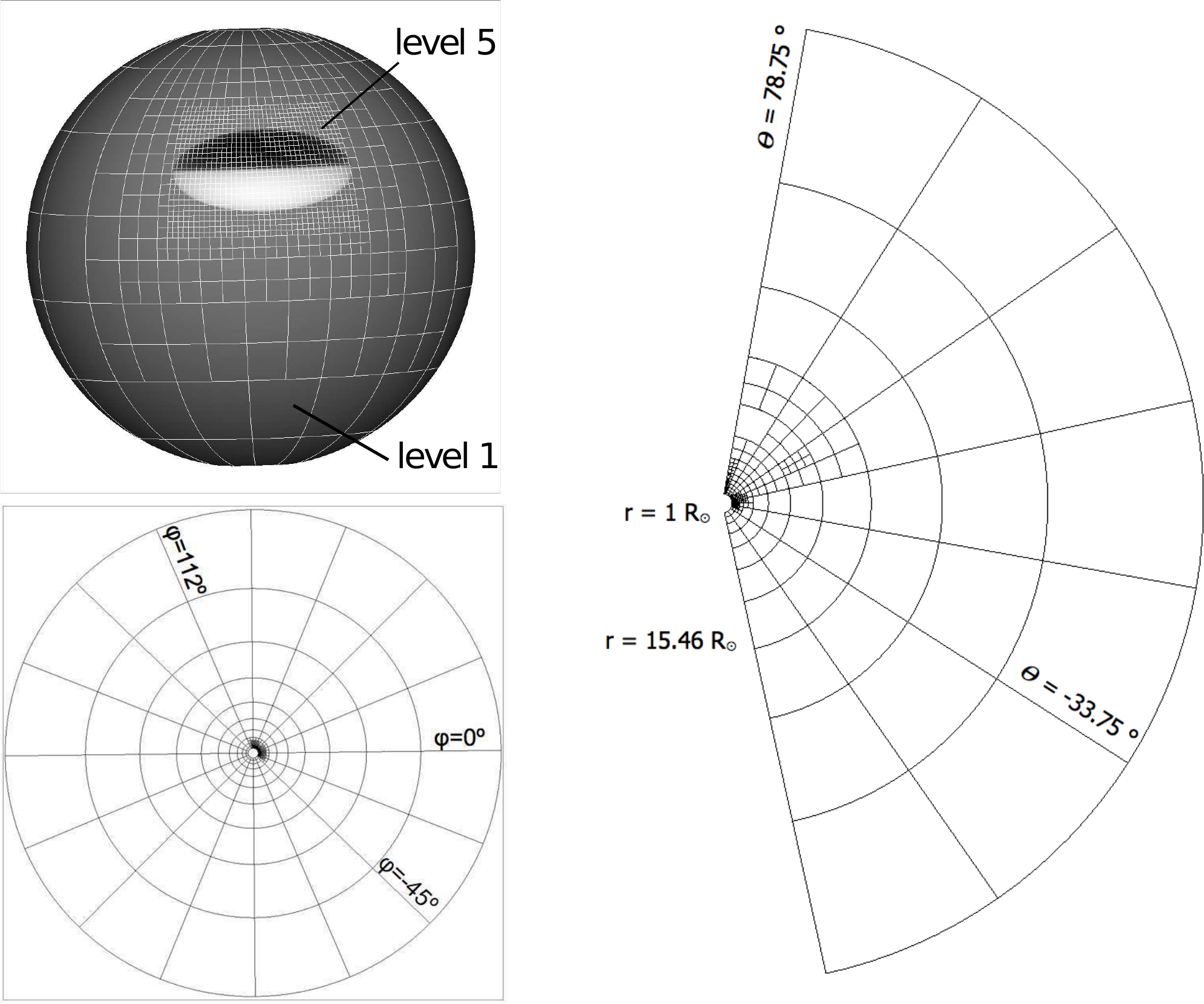}
 }
\caption{Grid properties: top left panel shows the grid imposed at the photosphere ($r = 1R_\odot$) at $ t= 0$ s. The region including the magnetic polarity cannot coarsen below level five. Left bottom and right panels show, respectively, the grid distribution in the [$r,\phi$] and [$r,\theta$] planes at an advanced time in the simulation ($ t= 5.74\times10^4$ s , i.e., $t_A= 74 $), when the adaptive mesh refinement is on.}
  \label{fig1}
\end{figure}

The increased number of mesh points in the 3D simulation ($30000\times8^3 \simeq15.36\times 10^6$) led us to restrict the domain where the grid is allowed to refine, in order to keep the computation time reasonable. 
The opening of the field under the solar wind pressure (\S\ref{atmosphere}) implies the formation of an extended heliospheric current sheet. While it is preferable to keep good resolution in the region where the field transitions between closed and open flux, it is not necessary further away.  In addition, the refinement is required only in the region where the dynamics develop. Therefore, the grid is allowed to refine only in the limited subdomain $r \in [1R_\odot,15.5 R_\odot]$, $\theta \in [11.75^\circ,123.75^\circ]$, and $\phi\in[-45.0^\circ,+112.5^\circ]$, up to level five. Everywhere else, we required that the grid maintain its coarsest possible resolution (Figure~\ref{fig1}). During the relaxation phase to reach a quasi-steady state, grid refinement is not needed. Thus, we turned on the mesh refinement at $t=4.4\times10^4~\rm{s}$, shortly before the relaxation phase ends (at $t=5\times10^4~\rm{s}$) and the photospheric forcing is applied (\S\ref{forcing}).

\subsection{Initial Atmosphere and Magnetic Field}
\label{atmosphere}

\begin{figure}
\centerline{
 \includegraphics[width=0.7\textwidth,clip=]{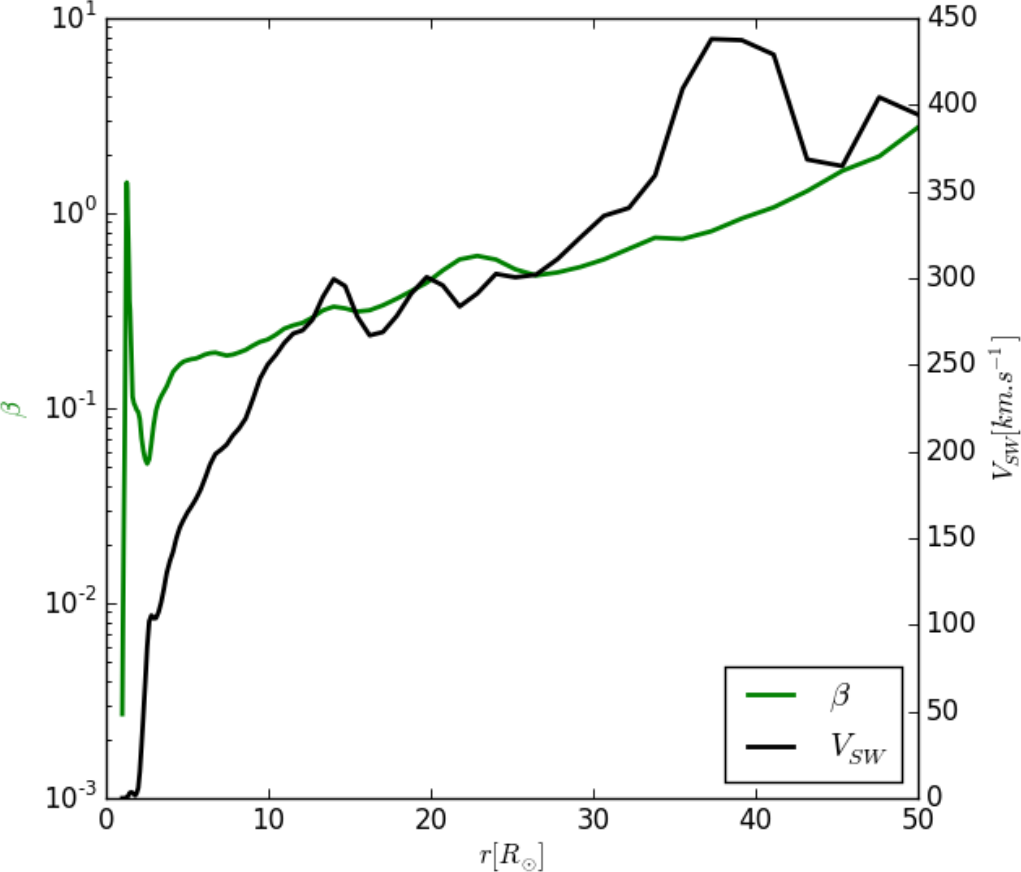}
 }
 \caption{ Radial profile of the plasma $\beta$ (green line) and solar wind speed $V_{SW}$ (black line) from the photospheric surface ($r = 1R_\odot$) to the outer simulation boundary ($r = 50R_\odot$). The 1D cut passes through the null point at the end of the relaxation phase ($t = 5\times10^4$ s, i.e. $t_A =0 $).}
   \label{fig2}
\end{figure}

The initialization of the atmosphere and the magnetic field has been adapted for the 3D geometry, but follows the same steps as in the 2.5D case. We initialized the plasma using the \citet{Parker58} model for a spherically symmetric, isothermal, trans-sonic solar wind,

\begin{eqnarray}
\frac{v^2}{c_s^2} \exp \left( 1 - \frac{v^2}{c_s^2} \right) = \frac{r_s^4}{r^4} \exp \left( 4 - 4\frac{r_s}{r} \right),
\label{eq-parker}
\end{eqnarray}
\noindent where $v(r)$ is the radial velocity, $c_s$ is the isothermal sound speed ($c_s^2 = 2 k_BT_0/m_p$) and $r_s = G M_{\odot}m_p/4k_BT_0$ is the radius of the sonic point. We assume a constant temperature $T_0=1 \times 10^6~\rm{K}$, for which $v = c_s \simeq 128$ km~s$^{-1}$ at $r = r_s = 5.8~R_\odot$. This yields an acceptable solar-wind speed of $420$ km~s$^{-1}$ at $50~R_\odot$. The inner-boundary mass density is a free parameter that we set to $\rho(R_\odot) = 3.05 \times 10^{-9}$ kg~m$^{-3}$, which yields realistic plasma $\beta$ values throughout the computational domain (see \S\ref{topology} and Figure \ref{fig2}).

\begin{figure}
\centerline{
 \includegraphics[width=0.9\textwidth,clip=]{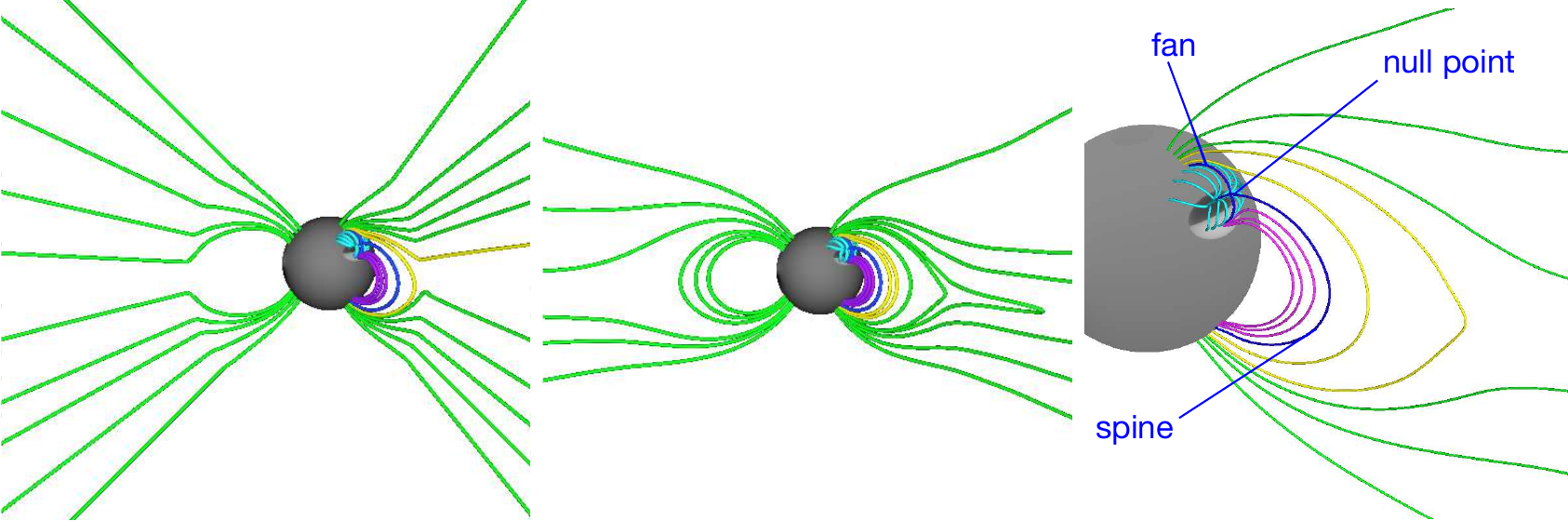}
 }
  \caption{Large-scale magnetic configuration between $0R_\odot$ and $20R_\odot$ before turning on the solar wind at $t = 0~\rm{s}$ (left) and after the system reaches a quasi-steady state at $t = 5\times10^4$ s, i.e.$t_A = 0$ (middle). The open and closed connectivities of the field are shown by the green magnetic field lines. Right panel: the null-point topology resulting from the inclusion of a dipolar active region in the north solar hemisphere. The photospheric radial magnetic field is grey-shaded with $B_r \in [-100, 100]$. Green field lines show open magnetic flux; yellow field lines show closed flux that belongs to the helmet streamer confining the null point. Dark blue field lines represent the fan and spine separatrices; light blue field lines show the fan shape of the flux enclosed below the fan/separatrix surface; and pink field lines show the flux closed below the outer spine and within the helmet streamer.}
\label{fig3}
\end{figure}

The velocity $v(r)$ computed from Parker's isothermal solution (Eq.\ \ref{eq-parker}), together with the associated mass density implied by the steady mass-flux condition $\rho v r^2 = {\rm constant}$,  defines the initial state of the plasma throughout the domain. We superimpose on that solution a potential magnetic field to create a quadrupolar magnetic configuration (\S\ref{topology}). The combined system initially is out of force balance, so first we relaxed it to a quasi-steady state. The 3D nature of the simulation increases tremendously the computational time. Since the relaxation phase is not the aim of our study, we sped up the relaxation by initially opening the large-scale magnetic field using the potential-field source-surface (PFSS) model for the background dipole (left panel in Figure~\ref{fig3}). After $t=5\times10^4~\rm{s}$, the kinetic and magnetic energies are essentially constant and no significant further evolution of the heliospheric plasma and magnetic field distributions is noted. Therefore, we considered that force balance between the outward solar-wind kinetic pressure and the inward magnetic-field tension was reached. The middle panel in Figure~\ref{fig3} shows the large-scale magnetic field at the end of the relaxation phase. The PFSS model over-estimated the amount of open magnetic flux, so that during the relaxation phase the field was closing down, i.e., the magnetic energy slowly increased. The quasi-steady magnetic energy at the end of the relaxation was higher than the minimum energy associated with the current-free state. This is due to the presence of a heliospheric current sheet, above the helmet streamer, which separates the open polar fields in the two hemispheres.

\subsection{3D Null-point Topology }
\label{topology}

The initial magnetic configuration emulates the basic ingredients
required for a solar eruption, in which a bipolar active region is
embedded in the global background magnetic field of the Sun. As in
MAD13, we positioned a point dipole at the center of the sphere to
create the background field, and we placed a simple bipolar active
region (AR) in the northern hemisphere. The sheared arcade that stores
the magnetic free energy required for the eruption is formed along the
polarity inversion line between the two polarities of the AR
\citep{DeVoreAntiochos08,Lynch_al08,Lynch_al09}. For this study, we
defined the active region by combining two half-elliptical magnetic
flux distributions, one of each polarity, at the photosphere, i.e.,
the inner boundary of the domain. The same procedure, with different
parameters, was used by \citet{DeVoreAntiochos08}. The surface radial
field distribution was decomposed into Fourier modes in longitude
$\phi$, and the Laplace equation for the magnetic scalar potential was
solved by matrix inversion in radius $r$ and colatitude $\theta$ for
each longitudinal mode. The Laplace solutions then were summed to
obtain the full potential magnetic field throughout the coronal
volume.

The location, orientation, and strength of the bipolar active region
were chosen to yield a multi-polar flux configuration with a 3D coronal
null-point \citep{Antiochos_al02,Torok_al09} and a closed outer spine,
which remained confined below the helmet streamer after the system
relaxed to its quasi-steady state. The positive (respectively
negative) flux distribution contained
$\Phi = +2.0 \times 10^{14}~\rm{Wb}$ ($-2.0 \times 10^{14}~\rm{Wb}$) with
the PIL centered at co-latitude $\theta_c = 66.5^{\circ}$ and
longitude $\phi_c = 28.5^{\circ}$. The polarity patches each 
have half-width $\Delta\theta_w = 11.5^{\circ}$ and half-length
$\Delta\phi_l = 20.0^{\circ}$. In this null-point topology, the
separatrix or fan surface passes through the null-point and delimits
the inner and outer connectivity domains. At the null, a singular
field line, the spine, intersects the fan surface. The inner spine is
anchored in the negative polarity below the fan; the outer spine is
anchored in the southern hemisphere. These topological objects are
labeled in the right panel of Figure~\ref{fig3} and illustrated by the
dark blue field lines. To visualise the surface of the fan, we plotted
light blue field lines that are enclosed below the dome-like
fan/separatrix surface. After the system relaxes, the outer spine
remains confined below the helmet streamer; the latter is illustrated
by the yellow lines in Figure~\ref{fig3}.

The prescribed background and active region magnetic fields provide a
realistic plasma $\beta$ over the whole numerical domain. The right
panel in Figure~\ref{fig2} shows a radial cut of $\beta$ from the
photosphere to $50R_\odot$, starting at the polarity inversion line
(PIL) of the embedded active-region dipole. One notices that
$\beta < 0.1$ in the corona $r=[0,4R_\odot]$, except near the coronal
null point where it sharply peaks to a large value ($\beta > 10$).  At
$r>4R_\odot$, in the interplanetary medium, $\beta$ begins a slow rise
that continues through the heliospheric domain up to the outer
boundary, but remains on the order of unity. Hence, our simulation is
performed in a solar/heliosphere-like regime.
 
\subsection{Boundary Forcing}
\label{forcing}

In the absence of any boundary forcing after the relaxation
  phase, our coronal system of magnetic field plus wind is in a fairly
  robust steady-state equilibrium, which simply oscillates mildly in
  response to small residual imbalances in the numerical magnetic,
  pressure, and gravitational forces. To obtain a CME eruption, we
  must energize our system by forming a filament channel. As in MAD13,
 this is done by imposing slow photospheric motions in the magnetic flux localized
  at the photospheric polarity inversion line.  In order to shear the magnetic
arcade along the PIL below the south lobe of the fan, in the $\phi$
direction, we prescribe two independent photospheric flows. The
line-tying boundary condition implies that a photospheric motion will,
in general, change the magnetic flux distribution at the solar
surface. In a 3D geometry, a shearing motion applied in only one
direction will induce some local accumulation of magnetic flux at the
surface, i.e., an increase of the Alfv\'en speed. Increasing the
  Alfv\'en speed decreases the numerical time step of the simulation
  in accord with the Courant-Friedrichs-Lewy stability condition. This
  may become critical if one wants to keep a reasonable computational
  time.

We defined a flow profile depending on the gradients of the magnetic field as given below, 
\begin{eqnarray}
\vec{v}(R_\odot,\theta,\phi) = v_0 \sin \left[ \pi \left( \frac{B_n-B_c}{B_r-B_l} \right) \right] \vec{e_r}\times\vec\nabla_t B_n(R_\odot,\theta,\phi).
\label{eq-flows}
\end{eqnarray}
The direction and magnitude of the flow are determined by the tangential gradient of the normal component of magnetic field, $B_n$. The flows are restricted to regions where $B_n$ ranges between $B_l$ and $B_r$. One of the chosen $B_l$, $B_r$ is near the extremal values at which $\vec\nabla_t B_n$ vanishes in the active region; the flow amplitude is made to smoothly decrease to zero at the other by appropriately setting the value of the parameter $B_c$. Outside the prescribed range, the flows are set to zero. The resulting circulation pattern preserves the surface contours of $B_n$ throughout the simulation.  The sign of the flow depends upon the sign of the magnetic field; the magnitude of the flow is set using a sine profile in order to avoid the formation of unresolved gradients at the flow boundaries. In the negative polarity we imposed the flow where $B_n$ is between $B_l = -49\times10^{-4}~\rm{T}$ and $B_r = B_c = -34\times10^{-4}~\rm{T}$, and in the positive polarity where $B_n$ is between $B_l = B_c = +36\times10^{-4}~\rm{T}$ and $B_r = +51\times10^{-4}~\rm{T}$. Because the magnetic flux distribution below the fan is not symmetric, the photospheric flows on either side of the PIL are also not symmetric. The resulting velocity field is shown in Figure~\ref{fig4}.

In the results given below, we express the time in Alfv\'en times rather than in seconds. This helps to characterize the dynamics of the system in terms of its characteristic time. The Alfv\'en time varies in the domain, so we chose to use the Alfv\'en time in the active region, which gives the characteristic time of propagation of information along the sheared arcade. The Alfv\'en speed in the active region at the surface is $v_A \simeq 2500$ km~s$^{-1}$ and the size of the active region is $L \simeq 250~\rm{Mm}$. These combine to define a characteristic Alfv\'en time $\tau_A=100~\rm{s}$.

The photospheric forcing is turned on at $t = 5\times10^4~\rm{s}$, after the relaxation phase. In the following, we consider this time as the new initial Alfv\'en time for the simulation, $t_A=0$. The unit of measure of the Alfv\'en time $t_A$ is the above-defined characteristic Alfv\'en time, $\tau_A= 100~\rm{s}$. The velocity fields are gradually applied using a temporal ramp by multiplying the spatial velocity field in Eq.\ \ref{eq-flows} with a time-dependent cosine function. The velocity plateaus at $ t_A=50 $, and is held fixed thereafter until time $ t_A=100 $. The maximum velocity ranges are $v_{phot}=14$--$35$ km~s$^{-1}$ in the positive polarity and $v_{phot}=13$--$48$ km~s$^{-1}$ in the negative polarity (Figure~\ref{fig4}). These speeds are greater than of the observed photospheric motion, but they are still less than $1\%$ of the Alfv\'en speed and less than half of the sound speed. We then ramped down the photospheric flows to turn them off at time $ t_A=112 $, soon after the CME was initiated.

\section{Null-point Reconnection}
\label{null point reconnection}

Figure~\ref{fig5} shows 2D cuts of the current density in the ($r,\theta$) plane at $\phi = 30^{\circ}$ at two times: $ t_A= 30 $ (left panel) and $ t_A= 52 $ (right panel). The black features show the intense electric currents (large $J$) while the white color shows no current ($J=0$). In both panels, there are three clusters of black arcs emanating from the photosphere. They highlight the volumetric current density along the sheared loop systems. The applied photospheric flows shear the loops anchored in the active region polarities (see \S\ref{forcing}). The left and right clusters of arcs correspond to the loops of the two side lobes of the null-point topology. The central cluster of arcs is the most intense and corresponds to the magnetic arcade sheared across the PIL, where the flux rope is expected to form through magnetic reconnection \citep{Aulanier_al10}.

As a result of the central-arcade shearing, electric current sheets form at the null point and along the fan/spine separatrices, above the central arcade and enclosed below the fan surface. At $t_A= 30 $, the null-point (NP) current sheet is weak and is only mildly elongated (Figure~\ref{fig5}, left panel).  At the later time $ t_A=52 $, as the forcing continues, the sheared arcades grow to redistribute the induced longitudinal ($\phi$) component of the magnetic field. The arcades push and compress the separatrices and tear apart the spines at the null point. This evolution extends and intensifies the NP current sheet and also strengthens its extensions along the separatrices \citep[Figure~\ref{fig5}, right panel; see, e.g.,][]{PonBatGal07a}.

In contrast to the simpler 2.5D case (MAD13), here the NP current sheet has a full 3D geometry. Figure~\ref{fig6} shows an isosurface of current-density magnitude. The field lines are colored the same as in Figure~\ref{fig5}, and so can be used as a reference to locate the current sheets with respect to the 3D null-point topology. Below the light blue field lines, we identify the volumetric current carried by the sheared arcade, and above them, the null-point current sheet.  The thin NP current sheet extends in both the $\theta$ and $\phi$ directions, and connects to both sides of the active region. This suggests that magnetic reconnection may occur not only in the vicinity of the null point, but anywhere along the whole NP current sheet.
 
The development of the NP current sheet makes it favorable for
magnetic reconnection to occur between the magnetic fields above and 
below the fan surface. To determine the onset of
reconnection at the null point, we looked for the first occurrence of
reconnection jets in the $(r,\theta)$ plane (not shown). The first
fast flows ($v_{flow} \simeq 130$ km~s$^{-1} = 2.7v_{phot}$) ejected
from the NP current sheet are noticed at $ t_A=44 $. 
Figure~\ref{fig7} presents the time evolution of the
connectivity of selected field lines once the null-point reconnection
is well established. The field lines are color-coded as in
Figures~\ref{fig5} and \ref{fig6}. Light blue field lines, enclosed
below the fan, reconnect at the null point with overlying yellow field
lines. As a result, the magnetic flux represented by the reconnected
yellow field lines is transferred below the null point on the north
side (see Figure~\ref{fig7}a,b,c,d). The flux represented by the
reconnected light blue field line jumps outside of the fan surface and
joins the outer-spine connectivity domain on the south side
(Figure~\ref{fig7}b,c,d). The flux transfer during this null-point
reconnection episode decreases both the magnetic flux confining the
null point (yellow field lines) and the flux overlying the sheared
arcade (light blue field lines). This null-point reconnection is
equivalent to the breakout reconnection that has been invoked to
explain the CME triggering in 2.5D simulations
\citep{Antiochos_al99,Karpen_al12}. The relationship between the onset
of breakout reconnection and the triggering of the CME has not been evaluated as
thoroughly in 3D breakout scenarios \citep{Lynch_al08,Lynch_al09}. In our simulation, the null-point reconnection ensures
that a similar flux transfer occurs. Whether that reconnection is the actual trigger of our CME, however,
is beyond the scope of the present study, and will be discussed in a
forthcoming paper. As a final note, in MAD13 we observed that the outer spine opened due to the null-point flux transfer prior to the onset of the eruption. In this 3D case, in contrast, the outer spine remains closed.

\section{Flux-Rope Formation}
\label{flux rope formation}

\subsection{Flare Reconnection}
\label{flare reconnection}

As described in \S\ref{null point reconnection}, the sheared arcades
carry volumetric currents and grow with time. In the 2.5D simulation
(MAD13), the growth of the arcades leads to the formation of a highly
extended current sheet, the so-called ``flare current sheet.'' Eventually
a stable X-point (in the 2D plane), accompanied by secondary X and O-points, 
is formed \citep{Karpen_al12,Guidoni_al16}. In 2.5D, the axisymmetry of the
system implies that flare reconnection results in the formation of a
CME flux rope that is disconnected from the solar surface and,
therefore, bounded by true separatrix surfaces. In 3D, however,
magnetic reconnection can occur in quasi-separatrix layers (QSLs) and,
generally, does not produce disconnected flux. QSLs are volumes where
the connectivity of the field changes rapidly, as measured by the
squashing degree $Q$ \citep{Demoulin_al96a, Titov_al02, Titov07}. 
QSLs are found surrounding separatrices, such as in null-point 
topologies \citep{Masson_al09b,Masson_al12a,
  Sun_al13,Pontin_al16,Masson_al17,Li_al18,Prasad_al18} and
pseudo-streamer topologies
\citep{Antiochos_al11,Titov_al11,Scott_al18}. They also can occur 
in the absence of null points and separatrices in bipolar regions
\citep{Mandrini_al91,Aulanier_al05,Aulanier_al06,Aulanier_al10}. 
Associated with the QSLs in a bipolar region, the hyperbolic flux tube
\citep[HFT;][]{Titov_al02} is the region where the QSLs are the
thinnest, i.e., the gradient of connectivity is the highest.  When a
flux rope is present, the HFT is located below the flux rope, and the
high Q volume separates the flux rope from the overlying field and the
post-flare loops \citep{Janvier_al13,Savcheva_al16,
  Masson_al17}. Electric currents develop in QSLs and reach their
maximum in the HFT where 3D magnetic reconnection can occur
\citep{Aulanier_al05,Aulanier_al06}.

The global magnetic configuration of our simulation consists of an
3D null-point topology confined below the helmet streamer
(yellow field lines in Figure~\ref{fig7}). However, the loop system
below the south lobe of the fan can be considered as a bipolar
magnetic configuration with a sheared arcade along the south section
of the PIL.  Figure~\ref{fig8} displays the time evolution of the 2D
cut in the $(r,\theta)$ plane of the electric current density at later
times in the simulation, after the sheared arcade has grown
significantly. At $ t_A= 84 $, the strong current-density structure in
the middle of the sheared arcade shows an Eiffel-tower shape, which is
the signature of the HFT in a vertical cut of the QSL
\citep[see][]{Janvier_al13}. 
The presence of such HFT suggests that magnetic reconnection can occur across the HFT \citep{Titov_al02, Aulanier_al06}. The black dome of current density below the HFT at $ t_A= 104 $ (Figure~\ref{fig8}, right panel) shows the current carried by the post-flare loops resulting from magnetic reconnection at the HFT.

Magnetic reconnection occurring across the HFT implies a continuous change of magnetic connectivity. This can be observed as an apparent slipping motion of the conjugate footpoints of field lines plotted from initial fixed footpoints \citep{Aulanier_al06}. The apparent slipping speed of the magnetic field lines is expected to be super-Alfv\'enic \citep{Masson_al12a,Janvier_al13}. To observe this slipping reconnection, we performed a new run of our simulation, producing output files every $t_A =0.5$, for a time interval ranging from $ t_A= 84 $ to $ t_A= 93.5 $. 
Figure~\ref{fig9} shows the time evolution of selected field lines. They are plotted from fixed footpoints in the positive polarity and are advected by the photospheric flows. By following the evolution of the conjugate footpoint of each field line, we find the slipping motion typical of 3D magnetic reconnection across the QSLs. Between$ t_A= 86.0 $ and $ t_A= 86.5 $, the footpoint of the green field line changes its connectivity from the middle of the active region to the left side of it (Figure~\ref{fig9}, top panel). A similar behaviour is noticed between$ t_A= 86.5 $, $ t_A= 87.0 $, and $ t_A= 87.5 $ for the dark blue field line (Figure~\ref{fig9}, three bottom panels), as well as for the other coloured field lines (see the online animation). The resulting loops of this magnetic reconnection across the HFT are of two types: (1) longer and more sheared field lines in the developing flux rope (see Figure~\ref{fig9}); and (2) shorter and less sheared field lines in the resulting post-flare arcade (not shown).

In Figure~\ref{fig9}, the vertical electric current density at the solar surface is greyscale-coded. Positive and negative electric current ribbons are observed on opposite sides of the PIL. The coloured field lines are anchored in and connect the two current ribbons. As time passes (see online animation), the reconnecting field line footpoints move along the negative (black) $J_z$ ribbon, supporting the idea that current ribbons correspond to the photospheric trace of the current-loaded QSLs \citep{Janvier_al16}.

The association of the rotational flows (\S\ref{forcing}) and the
magnetic reconnection of the sheared arcade at the HFT leads to the
formation of a flux rope. Whether the modeling is in 2.5D
\citep{Masson_al13} or in 3D \citep{Aulanier_al10,Zuccarello_al15},
the flux rope is formed by photospheric driving and magnetic
reconnection between the sheared arcades.
  
\subsection{Side Reconnection}
\label{Side reconnection}

In addition to the null-point reconnection (\S\ref{null point reconnection}) and flare reconnection (\S\ref{flare reconnection}), our system undergoes a third kind of reconnection. The longitudinal extension of the null-point current sheet along the fan (see Figure~\ref{fig6}) indicates that magnetic reconnection can occur not only at the apex of the fan dome, but also on the sides where intense currents develop  \citep{DeVoreAntiochos08,Lugaz_al11}. Following the connectivity of the flux-rope field lines, we identify that magnetic reconnection occurs on both sides of the fan, i.e. on the right and left of the active region.

In Figure~\ref{fig10}, we plotted four groups of coloured field lines according to their connectivity. Each field line is plotted from fixed footpoints in order to track the connectivity. Green and red field lines initially belong to the flux rope. The green field lines are traced from fixed footpoints in the positive polarity of the active region, while the red field lines are traced from fixed footpoints in the negative polarity. Yellow and blue field lines, respectively, connect the two solar hemispheres and are traced from fixed footpoints in the southern and northern hemispheres. Between $ t_A= 84 $ and $ t_A= 86 $, the red and blue field lines exchange their connectivities (Figure~\ref{fig10}~a and b). Initially localised in the photospheric flows inside the active region, a fraction of the flux rope (blue lines) is now rooted outside of the active region. Later, between $ t_A= 92 $ and $ t_A= 94 $, reconnection occurs on the other side of the fan between the blue and green flux-rope field lines and the yellow lines (Figure~\ref{fig10}c,d). As a result, the flux rope that initially is confined below the fan surface and represented by the green, blue and yellow field lines (Figure~\ref{fig10}d) now connects the two solar hemispheres.
In addition to the field lines, we display in Figure~\ref{fig10} a 2D cut of the current density in the $(r,\phi)$ plane. This shows that reconnection on the side of the fan occurs in regions of intense and thin current sheets (black structures).

The two side-reconnection episodes between the flux rope and the surrounding field modify the morphology of the flux rope. While the right (west) flux-rope footpoint is anchored in the positive polarity of the active region, the left (east) footpoint is partly anchored in the negative southern hemisphere (southeastern green, yellow, and blue lines in Figure~\ref{fig10}d) and partly in the negative polarity of the active region (northern green lines in Figure~\ref{fig10}d). Such reconnections with the surrounding field are likely to have changed the shear/twist of the CME.

Late in the buildup to solar eruption in our configuration, reconnection occurs between the active-region field and the surrounding background field. Such dynamics have been seen in other 3D MHD simulations, such as self-consistent flux-rope formation driven by slow photospheric flows \citep{DeVoreAntiochos08} and eruption of initially unstable flux ropes \citep{Lugaz_al11,VanDriel_al14}. Intense current sheets (greyshading in Figure~\ref{fig10}) that formed along the fan surface as a consequence of the photospheric forcing (see \S~\ref{null point reconnection}) are preferential sites for reconnection to occur. In principle, reconnection can occur anywhere along the fan surface; in our simulation, it occurs extensively along the sides of the active region, between the erupting flux rope and the surrounding magnetic flux. The location of the side reconnection is determined by the photospheric forcing, where the rotational flows intersect the fan footpoints: on the east and west extremities of the positive polarity. Thus, the flows move the fan/separatrix footpoints in the positive polarity. They deform the separatrix surface, increasing the magnetic shear in regions between the inner and outer fan fluxes. These regions on the east and west of the active region develop strong shear across the fan separatrix and intense current sheets, thereby becoming preferential sites for magnetic reconnection.

\section{Eruption}
\label{Eruption}

Figure~\ref{fig11} shows the evolution of the magnetic field of the CME represented by four groups of coloured field lines at six different times. Green, yellow, and red field lines are plotted from fixed footpoints in the negative and positive polarities of the active region; blue field lines are plotted from fixed footpoints anchored west of the active region. In the top panels (see also the online animation), the flux rope is forming as shown by the red and green field lines (see \S\ref{flare reconnection}). The flux rope then reconnects with the surrounding field on the right and on the left of the active region (respectively the top right and middle left panels of Figure~\ref{fig11}, and \S\ref{Side reconnection}). In the meantime, the flux rope is rising in the corona, first slowly and, after $ t_A= 82 $, more rapidly (see the online animation and the four bottom panels of Figure~\ref{fig11}). This evolution indicates that the eruption is underway.

To confirm the eruption, we plotted the time evolution of the magnetic and kinetic energies during the simulation (Figure~\ref{fig12}). Between $ t_A= 70 $ and $ t_A= 90 $, we identify a clear increase of the kinetic energy that is accompanied by the first decrease of the magnetic energy at $ t_A= 80 $. In 2.5D, the onset of the flare reconnection can be clearly determined \citep{Karpen_al12,Masson_al13,Guidoni_al16}. In 3D, the flare reconnection corresponds to magnetic reconnection at the hyperbolic flux tube (\S\ref{flare reconnection}). According to the 2D vertical cut of the current density (Figure~\ref{fig8}), the HFT is formed by $ t_A= 84 $ and the sheared arcade has started to reconnect at the HFT before $ t_A= 86 $ (Figure~\ref{fig8}). Therefore, we can assert that the flare reconnection starts before $ t_A= 86 $. While the timing is not precise, the onset of the flare reconnection corresponds temporally with the time interval of the kinetic energy increase. Such result has been found in 2.5D \citep{Karpen_al12}, and our detailed analysis suggests that such temporal association may be generalised to 3D. However, whether the flare reconnection triggers the impulsive increase of the kinetic energy requires a thorough study that is not the aim of this paper. We do not observe a clear decrease in the magnetic energy at this time because the photospheric flows are applied steadily until $ t_A= 100 $ and then ramped down to zero at $ t_A= 112 $. After that, the magnetic energy levels off and the kinetic energy decreases steeply as the reconnection that is driven at the separatrix and within the flare sheet by the footpoint motion quickly subsides. 


The slow and impulsive rising phases obtained from the global dynamics of the field (see online material) are confirmed by computing the velocity of the flux rope. Finding the apex of the flux rope in a 3D asymmetric system at one particular time can be done easily \citep{Zuccarello_al15}, but following it in time is not straightforward.  For simplicity and clarity, we chose to follow the apex of the magnetic arcade located above the flux rope but below the null point. This permits us to follow the rising phase of the system before the flux rope forms. Although those loops are not sheared, they are pushed and driven by the growing sheared loops underneath.  Once the system enters the impulsive phase, the overlying loops are either part of the flux rope (due to flare reconnection) or have been removed through null-point reconnection. To determine the location of the apex of those arcades, we searched for the local minimum of the radial component of the magnetic field along a 1D radial cut at $(\theta\phi) = (23^{\circ},30^{\circ})$ (the center of the straight segment of the PIL). Then we extracted the radial speed of the plasma at this location. The right panel of Figure~\ref{fig12} shows the temporal evolution of the radial velocity. Before the null-point reconnection starts ($ t_A= 44 $), the arcade hardly rises at all. Once magnetic flux reconnects at the null point, we notice a small and constant rise of the structure until $ t_A= 75 $. Then a change in the radial speed evolution occurs, and the flux rope accelerates, reaches a maximum speed of $v_{cme} \simeq 1400$ km~s$^{-1}$, and then plateaus at $v_{cme} \simeq 900$ km~s$^{-1}$. Since the CME speed is extracted from a single radial cut, the maximum may not be representative of the global CME speed, but it provides an indication of the kinematics of the eruption. Examining 2D cuts in the ($r,\theta$) and ($\theta,\phi$) planes (not shown), we find small-scale and localised structures that move faster than $1400$ km~s$^{-1}$. We estimate the global maximum speed of the CME to be in the range $v_{cme} \simeq 1000$--$1200$ km~s$^{-1}$. 
The change of behaviour of the radial speed near $ t_A= 75 $ is approximately temporally coincident with the increase of the kinetic energy and the onset of the flare reconnection at the HFT.

\section{Coupling the CME to the Interplanetary Magnetic Field}
\label{Coupling the CME and the interplanetary magnetic field}

After the eruption, the CME propagates through the high corona into
the interplanetary medium, where it interacts with the ambient
magnetic field. Figure~\ref{fig13} shows the temporal evolution of the
CME magnetic field during its propagation into the high
corona/interplanetary medium. Figure~\ref{fig13}a shows the initial
field before any interaction of the CME (blue and green field lines)
with the interplanetary magnetic field (yellow field lines). The blue,
green, and yellow field lines are plotted from fixed footpoints at the
solar surface. Any changes in the connectivity of those magnetic field
lines are therefore caused by magnetic reconnection. In
Figure~\ref{fig13}b some initially closed blue and green field lines
have opened into the interplanetary medium, indicating that the CME
has reconnected with the open interplanetary field close to the Sun in
our 3D geometry.  As the CME propagates further into the
interplanetary medium, more magnetic flux of the CME reconnects with
the open field of the interplanetary medium
(Figure~\ref{fig13}c,d,e,f; see also the online material).

During the flux-rope buildup phase, the blue field lines are brought
into the forming flux rope through magnetic reconnection with the red
sheared arcade (\S\ref{Side reconnection}, Figure~\ref{fig10}).  As
shown in Figure~\ref{fig11} (see online material), those blue field
lines initially belong to the core of the flux rope and are plotted
from the same footpoints as those reconnecting with the open field in
Figure~\ref{fig13}. In contrast to MAD13, where the magnetic
reconnection progresses from the external field to the core field, in
this 3D geometry the open field is able to reach into and reconnect
with the core field of the flux rope, producing complex ejecta
entangling open and closed field. A critically important difference
between these results and those in MAD13 is that, contrary to the 2.5D
case, not all of the erupting flux rope reconnects with the open
field. A clearly distinguishable flux rope survives to propagate out
to the heliosphere. This result agrees with in-situ measurements
showing that flux-rope CMEs associated with impulsive SEPs reach the
Earth \citep{Wimmer_al06}.


\section{Application to the Escape of Flare-accelerated Particles}
\label{Application to the escape of flare-accelerated particles}

Energetic particles in the corona have a small Larmor radius and closely follow the reconnected magnetic field lines.  Therefore, the dynamics of the magnetic field during reconnection can be tracked to approximate the injection channel of accelerated particles \citep[e.g.][]{Aulanier_al07,Masson_al09b,RosGals10}. Electromagnetic emissions (UV, X-rays, $\gamma$-rays), generated by particles impacting the denser chromospheric layer, are localized at the footpoints of the reconnected field lines \citep[see, e.g.][]{Demoulin_al97,Masson_al09b,Reid_al12,Musset_al15,Masson_al17}. According to the standard 3D model of eruptive flares, solar particles accelerated at the flare current sheet behind the CME should remain trapped, either in the post-flare loops or on the CME field lines \citep{Reames13}. By modeling a 3D  self-consistent solar eruption, we show that after the CME takes off, it reconnects with the open interplanetary magnetic field. The newly reconnected flux couples the open IMF and the CME, providing a magnetic path for flare-accelerated particles, initially confined in the CME, to escape onto the open field. This 3D simulation verifies our model for flare-accelerated particle escape, which was tested initially in 2.5D. Moreover, the full 3D geometry of our new simulated eruption allows us to determine the spatial and temporal properties of the escaping particle beams.

Particle acceleration during the flare relies on small-scale processes
developing in the current sheet where magnetic reconnection develops
\citep{Cargill_al12}. In our 3D CME model, we identified three reconnection sites
as potential acceleration sites for solar particles: the HFT current
sheet (Figure~\ref{fig8}), and the two current sheets along the fan
where the side-reconnection episodes occur
(Figure~\ref{fig10}). Particles that have been accelerated in those
current sheets would be injected onto the newly reconnected field
lines and, therefore, populate the forming flux rope. Going beyond the
2.5D model, our 3D results indicate that particles accelerated at the
flare current sheet behind the CME subsequently can be further accelerated
at the current sheets at the side-reconnection sites
(\S\ref{Side reconnection}).

Starting at a radial distance of $5{R}_\odot$ and sweeping over
$\theta \in [0,60]^\circ$ and $\phi \in [0,60]^\circ$, we plotted
field lines back to the solar surface in order to estimate the
longitudinal and latitudinal distribution of the CME/open-IMF flux. By
selecting field lines that couple the CME and the open IMF, we
estimate the longitudinal extension to be
$\Delta \phi \simeq 30^\circ$ and the latitudinal extension to be
$\Delta \theta \simeq 6^\circ$. Dark blue field lines in
Figure~\ref{fig13}e,f show the longitudinal broadening of the CME/IMF
flux. This longitudinal extension corresponds roughly to the width of
the CME front. Although a dedicated study is required to determine
whether the width of the CME is a limiting factor for the spatial
extension of the new CME/IMF flux, our 3D model suggests that
flare-accelerated particles can be injected over a longitudinal range
comparable to the width of the CME.

During the eruption, the flare and side reconnections build up the
flux rope (\S\ref{flare reconnection}), then the CME reconnects with
the interplanetary field (\S\ref{Coupling the CME and the
  interplanetary magnetic field}). This sequential order in the
reconnection episodes suggests that the acceleration and the escape of
energetic particles are separated in time. Determining accurately the
onset of 3D magnetic reconnection is very challenging. Nonetheless, we
can estimate an approximate onset time by combining several
indicators. The development of an intense current sheet at the HFT
indicates that magnetic reconnection can take place. Tracing many
field lines will also help to identify when magnetic reconnection
starts; thus, for the flare reconnection, we
estimate the onset time around $t_{A,acc} \simeq 84 $ (\S\ref{flare
  reconnection}). Based on the field-line connectivity changes, the
side reconnection on the right of the active region is estimated to
start at $t_{A} \simeq 86 $, and the reconnection between the open field
and the CME starts at $t_{A,esc} \simeq 104 $ (Figure~\ref{fig11},
bottom right panel). Thus, assuming that the first particles to be
accelerated at the flare are also the first particles to escape, the
lower limit of the time interval between the acceleration and the
escape is $\Delta t_{A,sim} = t_{A,esc}-t_{A,acc} \simeq 20$. 
In absolute units, this time interval is $\Delta t_{sim} = 
\Delta t_{A,sim} \times \tau_{A,sim} \simeq 2000~\rm{s} \simeq 33~\rm{min}$.

As in MAD13, we can scale our simulation to an observed event. 
The Alfv\'en time for our modeled active region is $\tau_{A,sim}$$ = 100~\rm{s}$
(\S\ref{forcing}). For a young and strong active region, a typical
length can be $L_{obs} \simeq 50~\rm{Mm}$ with a maximum Alfv\'en speed
$c_{A,obs} \simeq 5000$ km~s$^{-1}$, giving a minimum Alfv\'en time 
$\tau_{A,obs}$$ = 10~\rm{s}$. The conversion factor between the
simulated and observed active regions is $\tau_{A,sim} = 
10~\tau_{A,obs}$ which yields a time interval 
$\Delta t_{obs} = \Delta t_{sim} / 10 = 200~\rm{s}$. 
Hence, the time interval between the acceleration and the injection of 
particles is at least $\Delta t_{obs} \simeq 3~\rm{min}$, but it can 
be larger. Applying the same scaling factor to the time interval for
particles accelerated at the onset of the flare reconnection but
escaping later during the CME/open-field reconnection episode -- e.g.,
$t_{A,esc} = 128 $ as in Figure~\ref{fig13}f -- we obtain a time
interval $\Delta t_{obs} = 440~\rm{s} \simeq 7~\rm{min}$. We have used
a high conversion factor based on a young, compact active region. For
an active region larger by a factor of two, the conversion factor
decreases by the same factor and the minimum and maximum time
intervals double, to $\Delta t_{obs} \simeq 6~\rm{min}$ and
$14~\rm{min}$, respectively. Therefore, contrary to MAD13, where we
suggested a prompt injection of the flare-accelerated particles with
$\Delta t_{obs} < 1~\rm{min}$, our 3D results suggest that particles
can be trapped in the ejected flux rope for a few minutes before
accessing the open field.

\section{Implications for Observations}
\label{Discussion}

For this paper, we performed a 3D MHD simulation in spherical
coordinates of a CME eruption from a null-point topology,
embedded in a helmet streamer whose boundary is defined by an
isothermal solar wind. The goal was to generalise and extend our model
for the escape of flare-accelerated particles initially developed in
2.5D \citep{Masson_al13}.  Our study provides a detailed analysis of
the dynamics of a CME erupting into a background solar wind. The
initial magnetic topology is an asymmetric null-point topology with a
closed outer spine that is confined below the helmet streamer.  During
the initiation of the CME, the energy builds up by due to the
application of slow rotational flows in the active-region polarities
below the null point. This shears the magnetic arcade above the
straight part of the PIL separating the polarities. Magnetic energy
release occurs through null-point reconnection first, followed by
flare reconnection that forms the flux rope. The 3D geometry, the
null-point topology, and the presence of a large-scale background
magnetic field cause additional reconnection episodes to develop on
the flank of the fan surface between the flux rope and the surrounding
field. Those side-reconnection episodes change the global morphology
of the flux rope. When the balance of forces in the system fails, the
flux rope erupts into the corona and propagates into the
interplanetary medium. Connections between the CME field lines and the
open interplanetary magnetic field lines are established via
interchange reconnection during the propagation of the CME. The extension of our
model to 3D allows us to derive temporal and spatial properties of
flare-accelerated particle beams that may be generated by the
eruption.

In our model, the flux rope builds up as the result of 3D magnetic
reconnection across a hyperbolic flux tube. This process has already
been shown explicitly for one 3D MHD zero-$\beta$ simulation of an
erupting flux rope \citep{Aulanier_al10,Janvier_al13}; our study
confirms it for a nonzero-$\beta$ system. Using ARMS, several 3D MHD
simulations of CME eruptions
\citep{DeVoreAntiochos08,Lynch_al08,Lynch_al09,Dahlin_al19} have been performed in
which 3D magnetic reconnection occurred without a true separatrix. The
formation and reconnection of an HFT may have been responsible for the
flux rope formation during those simulated eruptions, as well, but
that aspect of the evolution was not investigated or reported.

To our knowledge, this is the first nonzero-$\beta$ 3D MHD simulation
self-consistently producing a fast CME ejected into a solar wind as a
result of simple photospheric shearing flows. In
our model, the flux rope is not inserted ab initio, but instead builds
up dynamically as a result of applied footpoint shear and magnetic
reconnection. The values of magnetic field and plasma pressure have
been chosen in order to produce a simulation with a realistic plasma
$\beta$, i.e., $\beta \ll 1$ in the corona and $0.1 \le \beta \le 1$
beyond $r = 5R_\odot$. The CME speed reaches $v_{cme} \simeq 900$
km~s$^{-1}$, which is typical for fast CMEs. Other numerical studies
have modeled CME triggering and eruption within other
scenarios. \cite{Lynch_al08} obtained a fast ($v_{cme} \simeq 1200$
km~s$^{-1}$) self-consistent breakout CME in a corona with a realistic
$\beta$ but a static atmosphere rather than a
wind. \cite{Aulanier_al10} and \cite{Zuccarello_al12} also modeled
self-consistent CMEs, but the speeds of the ejecta were smaller,
$v_{cme} \simeq 400$ km~s$^{-1}$, from their zero- and nonzero-$\beta$
atmospheres, respectively.

Our global modeling of the system, including the large-scale background magnetic field, allows magnetic reconnection between the active-region and surrounding magnetic fields to develop. After the flare reconnection, the forming flux rope reconnects at the extended null-point current sheet with magnetic flux to the sides of the fan (\S\ref{Side reconnection}). Besides building up the flux rope further, those side reconnections provide potential sites for particle acceleration. Thus, our results highlight that energetic particles may experience at least one acceleration episode in addition to that associated with the flare reconnection. As described in \S\ref{Application to the escape of flare-accelerated particles}, particles accelerated multiple times at different current sheets may populate the erupting flux rope.

Using radio imaging of a CME, \cite{Bain_al14} showed that flare-accelerated electrons can be trapped for hours within a CME. This observation supports the conclusion from our modeling that energetic particles produced by the flare are injected onto the flux-rope field lines. It is common to observe a time delay between the flare radiative signatures -- X-ray and EUV emissions indicating particle acceleration -- and the release of energetic particles into space, as inferred from SEP in-situ measurements assuming a Parker-spiral IMF configuration. This delay is often attributed to shock acceleration, which needs time to accelerate particles after the flare impulsive phase has concluded. Trapping particles in the CME is an alternative explanation for those delays \citep{Kocharov_al17}. Our model yields a delay of about ten minutes between the times of acceleration by flare reconnection and of escape along the open field by CME/IMF reconnection. This result confirms that particle trapping can be an alternative to shock acceleration for explaining the delayed arrival of energetic particles. We point out that radio observations also support the release of energetic particles through magnetic reconnection between the CME and the open IMF \citep[see][\S8, for discussion]{Masson_al13}.

With the STEREO spacecraft, it has been possible to detect and study the so-called widespread SEP events \citep{Dresing_al12,Lario_al13}. Both shock acceleration \citep{Richardson_al14,Lario_al14} and strong perpendicular diffusion \citep{Dresing_al12,Droege_al14,Strauss_al17} during the propagation in the IMF can explain characteristics of the longitudinal extent of the particle beams. Recently, \cite{Dresing_al18} suggested that the late release of CME-trapped flare-accelerated particles can explain how the particle fluxes of a widespread SEP event have the same intensity at two spacecraft separated by $60^\circ$. According to their detailed study, shock acceleration cannot explain the observations, but a sudden opening of the CME trap could do so. Our 3D model provides the physical mechanism that explains how CME closed field can open suddenly, releasing trapped energetic particles. Moreover, the longitudinal extent of the opening attains the whole width of the CME front, $\Delta \phi \simeq 30^\circ$. While this angle is not large, it is likely to depend upon the width of the CME; a wider CME would be expected to have a wider particle-release channel. In our model, the large-scale open background field expands radially. However, the open field in the corona can expand faster than $1/r^2$ as shown by observations \citep[e.g.][]{Klein_al08} and by modeling \citep{Antiochos_al11,Higginson_al17} for more complex background surface flux distributions. Reconnection between a CME and a highly diverging field would be expected to increase the spatial extent of the escaping particle beams.


\section{Conclusion}
\label{Conclusion}
The primary result of our 3D simulations is that the magnetic
reconfiguration of the corona through multiple reconnection episodes
can explain many of the observed properties of SEP events. In
particular, we show that the release of flare-accelerated particles
through the opening via reconnection of the CME flux-rope field is
sufficient to explain observed delays between the flare and the
arrival of SEPs at 1 AU and their wide longitudinal spreads. Our 3D
MHD model of reconnection-enabled particle escape presents a
compelling alternative to the CME-driven shock scenario commonly used
to explain the delays and the spatial distribution and location of the
SEPs.  \cite{Klein_al11} showed that western CME-less flares,
accelerating particles in the corona, do not produce any SEP event at
the Earth. Our study supports their statement: a CME is not required
in order for particles to be accelerated, but it may be essential if
those energetic particles are to escape the Sun promptly and travel to
Earth. Finally, and most important, we find that in 3D, energetic
particles can escape onto open field lines from deep within the CME
flux rope, and yet the flux rope survives as a distinct structure in
interplanetary space. Further work, including detailed event studies
now needs to be done in order to validate the model against particular
impulsive SEP observations.

We thank our anonymous referee for his/her comments that helped 
us to improve the manuscript. SKA and CRD acknowledge the support 
of NASA's LWS TR\&T and H-ISFM
research programs. The numerical simulations were carried out, in
part, using resources provided by NASA's HEC program on the Discover
cluster at the NASA Center for Climate Simulation.
\bibliographystyle{apj}


\begin{figure}
\centerline{
 \includegraphics[width=0.9\textwidth,clip=]{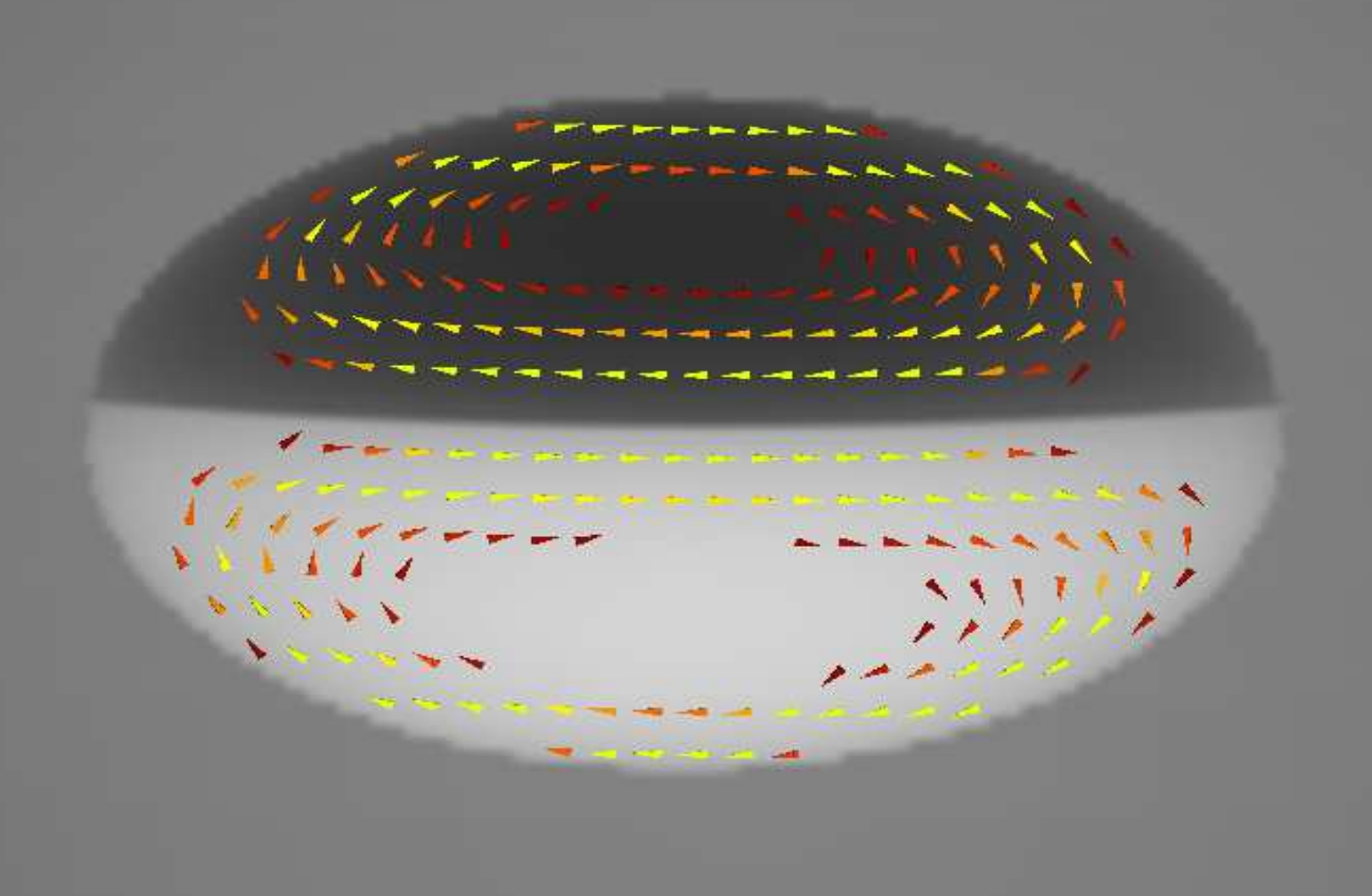}
 }
 \caption{ Photospheric flows: A zoom on the active region showing the grey-shaded radial magnetic field at the photosphere. The photospheric flows applied in the bipolar active region are shown as arrow heads colored by velocity value from $v^{min}_{phot} \simeq 0$ km~s$^{-1}$ (red) at the the periphery of the flow patterns to $v^{max}_{phot} \simeq 50$ km~s$^{-1} < 1\%~\rm{c_A}$ (yellow) in the center.}
\label{fig4}
 \end{figure}

\begin{figure}
\centerline{
 \includegraphics[width=0.9\textwidth,clip=]{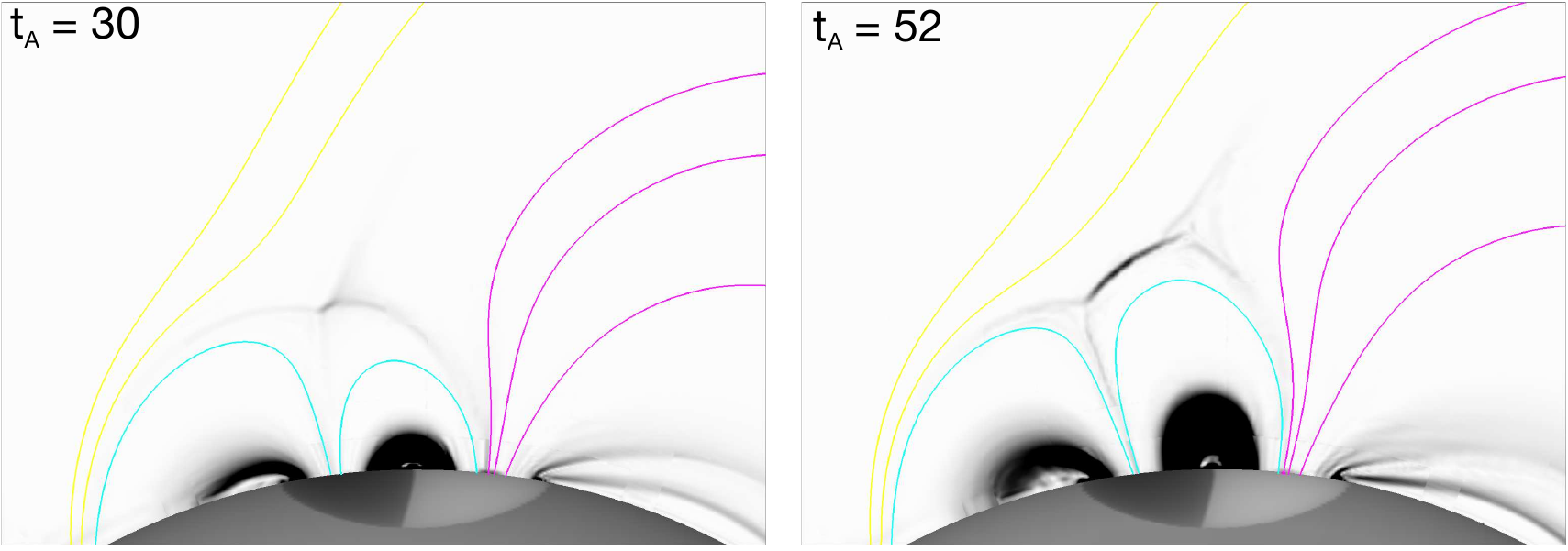}
 }
 \caption{Null-point current sheet. 2D cuts of the current density
   (sky plane) and the radial magnetic field (solar surface) for the
   null-point topology after the onset of the photospheric forcing, at
   times $ t_A=30 $ and $ t_A=52 $. The greyscale filled contours show the radial
   magnetic field at the photosphere (surface at the bottom of each image)
   and the current density in the $(r,\theta)$ plane at
   $\phi = 30^\circ$ (longitudinal center of the active region). The
   field lines are colored as in Figure~\ref{fig3}. Light blue and
   pink lines show, respectively, the magnetic field closed below the
   fan separatrix and closed below the outer spine; yellow lines
   belong to the helmet streamer confining the null point.}
  \label{fig5}
\end{figure}

\begin{figure}
\centerline{
 \includegraphics[width=0.9\textwidth,clip=]{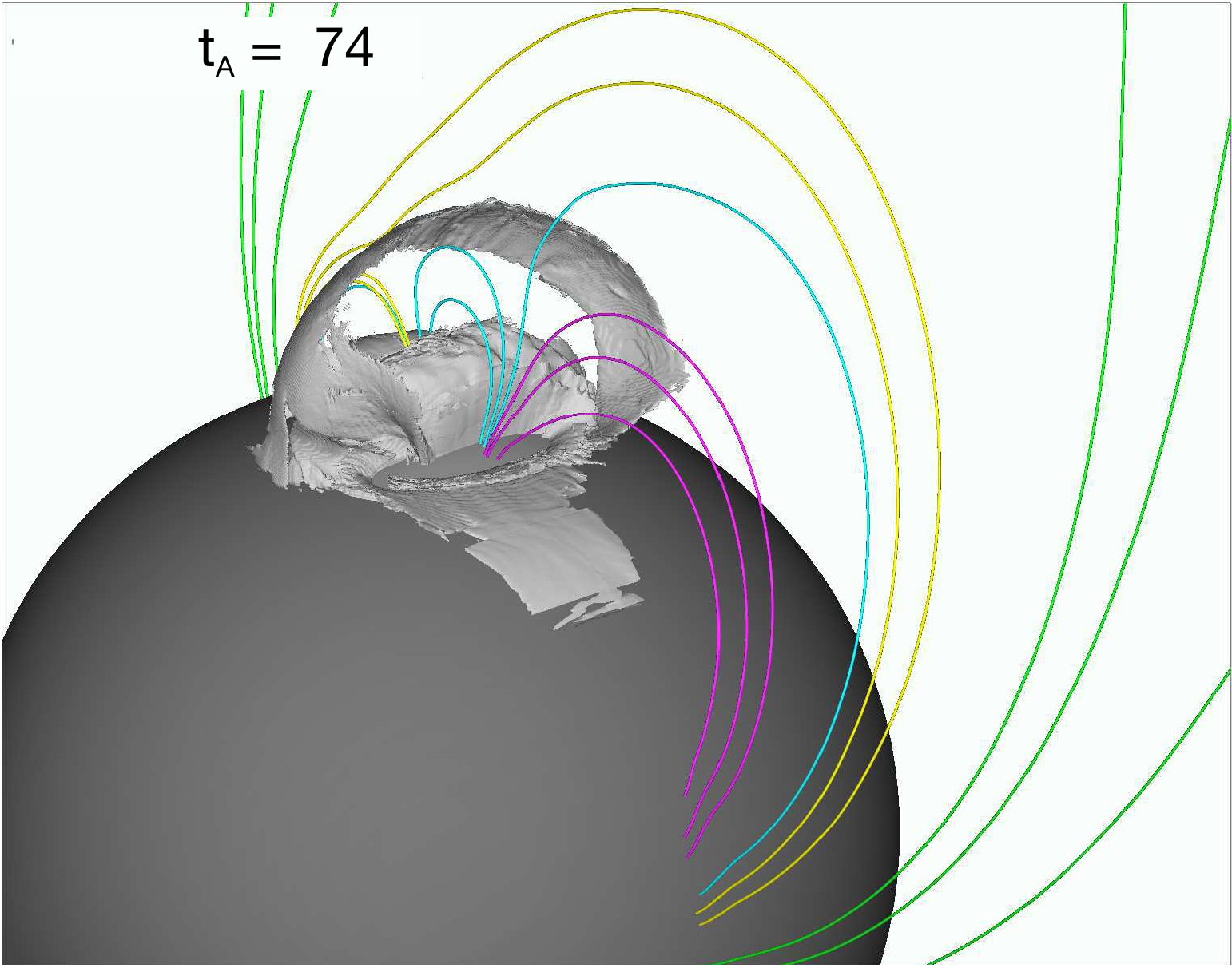}
 }
\caption{3D view of the null-point current sheet. The grey isosurface shows the current density magnitude $J=2\times10^4$ A m$^{-2}$. Magnetic field lines show the connectivity using the same color code as in Figures \ref{fig3} and \ref{fig5}.}
  \label{fig6}
\end{figure}

\begin{figure}
\centerline{
 \includegraphics[width=0.9\textwidth,clip=]{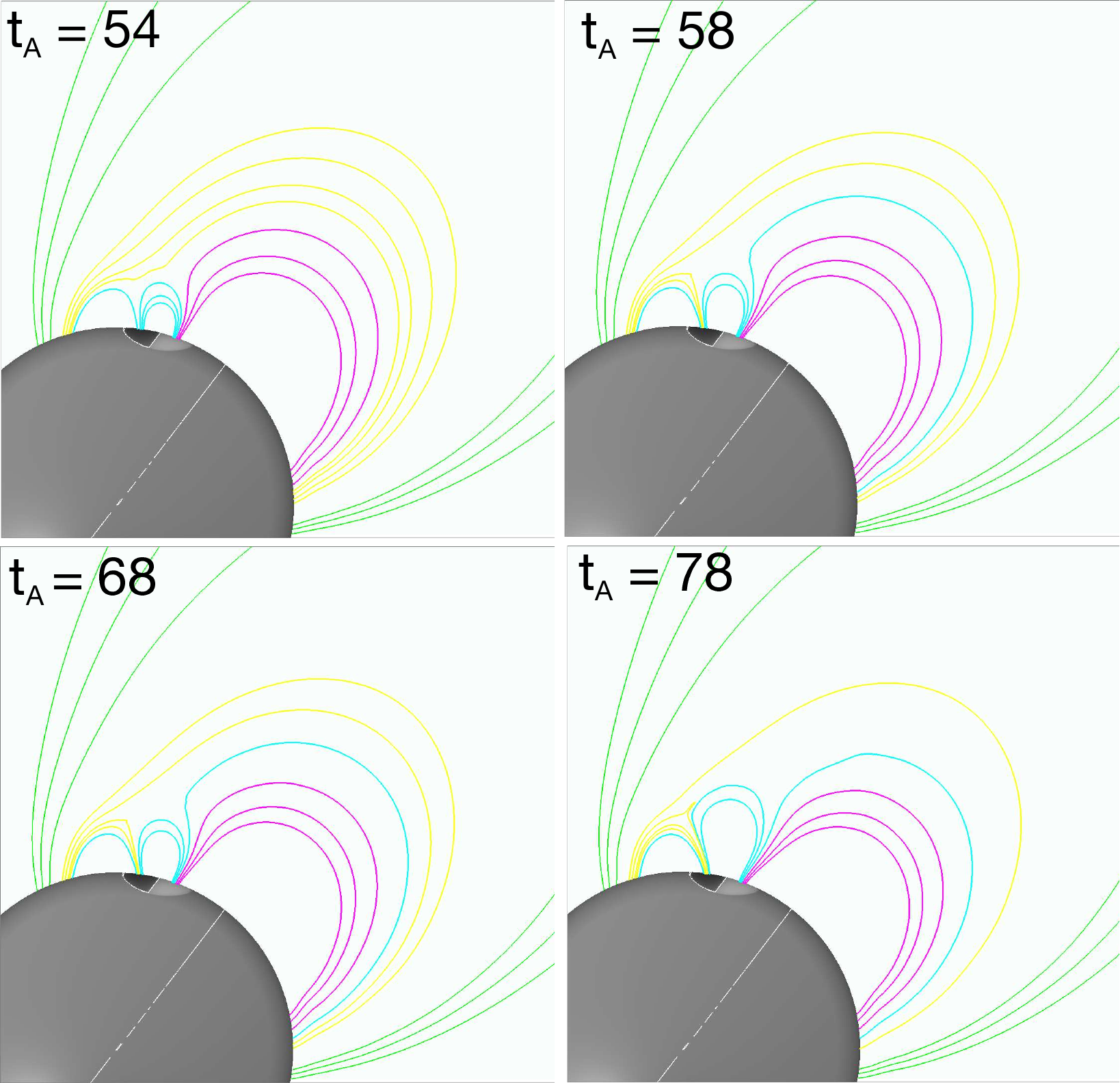}
 }
\caption{Connectivity evolution of the magnetic field during the null-point reconnection. The four panels show different times after the onset of null-point reconnection. Field lines are color-coded as in Figures \ref{fig3} and \ref{fig5}. See text for details.}
  \label{fig7}
\end{figure}

\begin{figure}
\centerline{
 \includegraphics[width=0.9\textwidth,clip=]{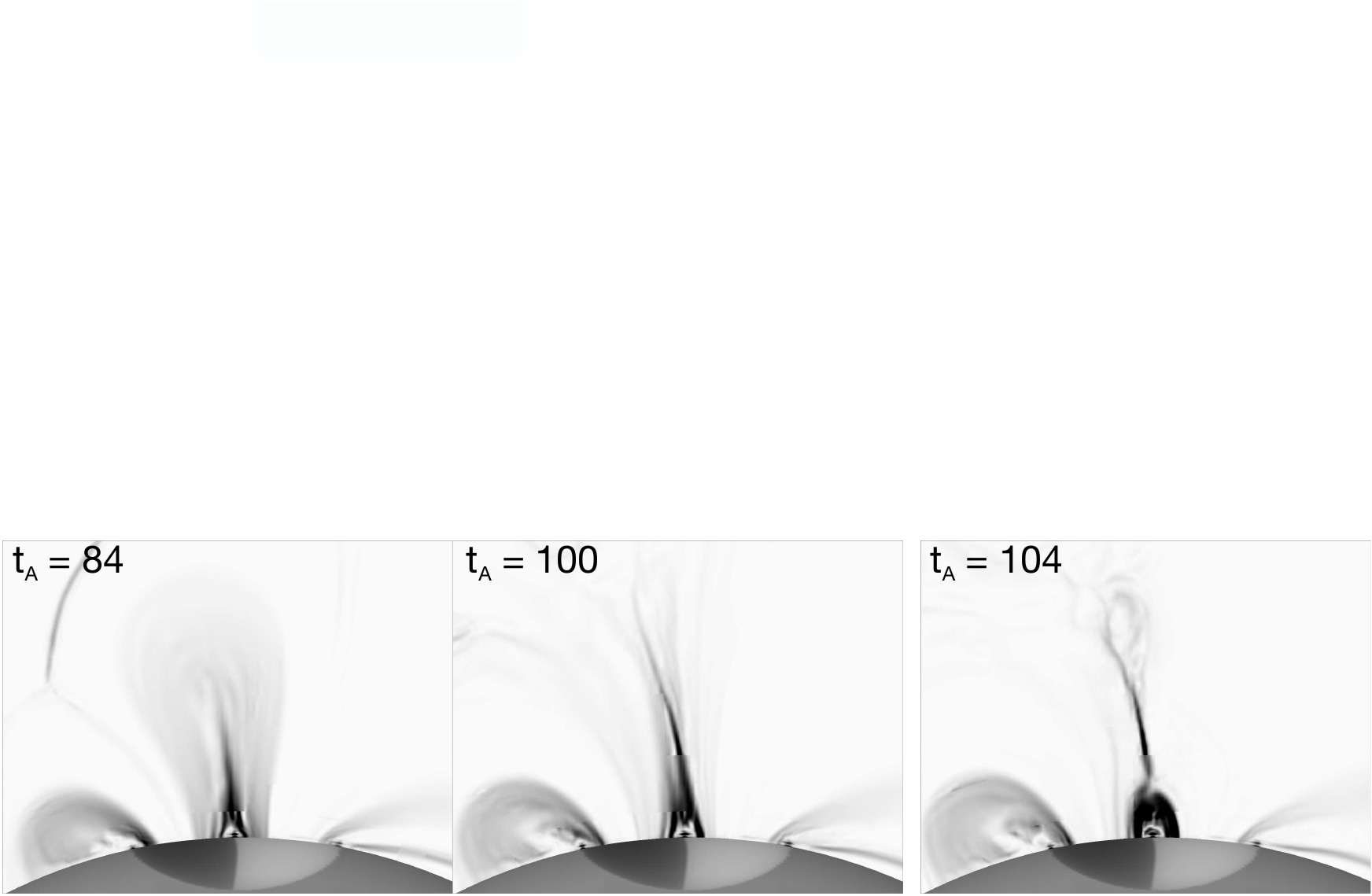}
 }
 \caption{Current density in the hyperbolic flux tube. The three
   panels show 2D cuts of the greyscale-coded current density in the
   $(r,\theta)$ plane at $\phi = 30^\circ$ at different times. The
   surface at the bottom of each panel displays the radial magnetic
   field in greyscale.}
  \label{fig8}
\end{figure}

\begin{figure}
\centerline{
 \includegraphics[width=0.3\textwidth,clip=]{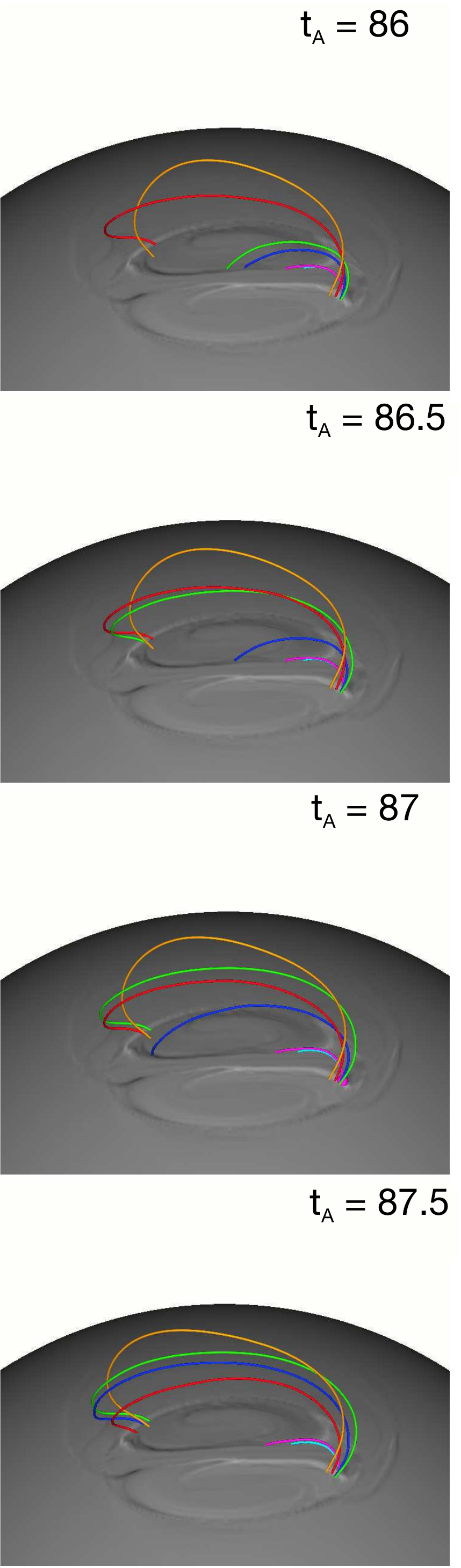}
 }
\caption{Reconnection inside the hyperbolic flux tube. Each coloured field line is plotted from fixed footpoints advected by the photospheric flows in the positive south polarity. The vertical current density is grey shaded at the photospheric boundary. An animation of this figure is provided online.}
  \label{fig9}
\end{figure}

\begin{figure}
\centerline{
 \includegraphics[width=0.9\textwidth,clip=]{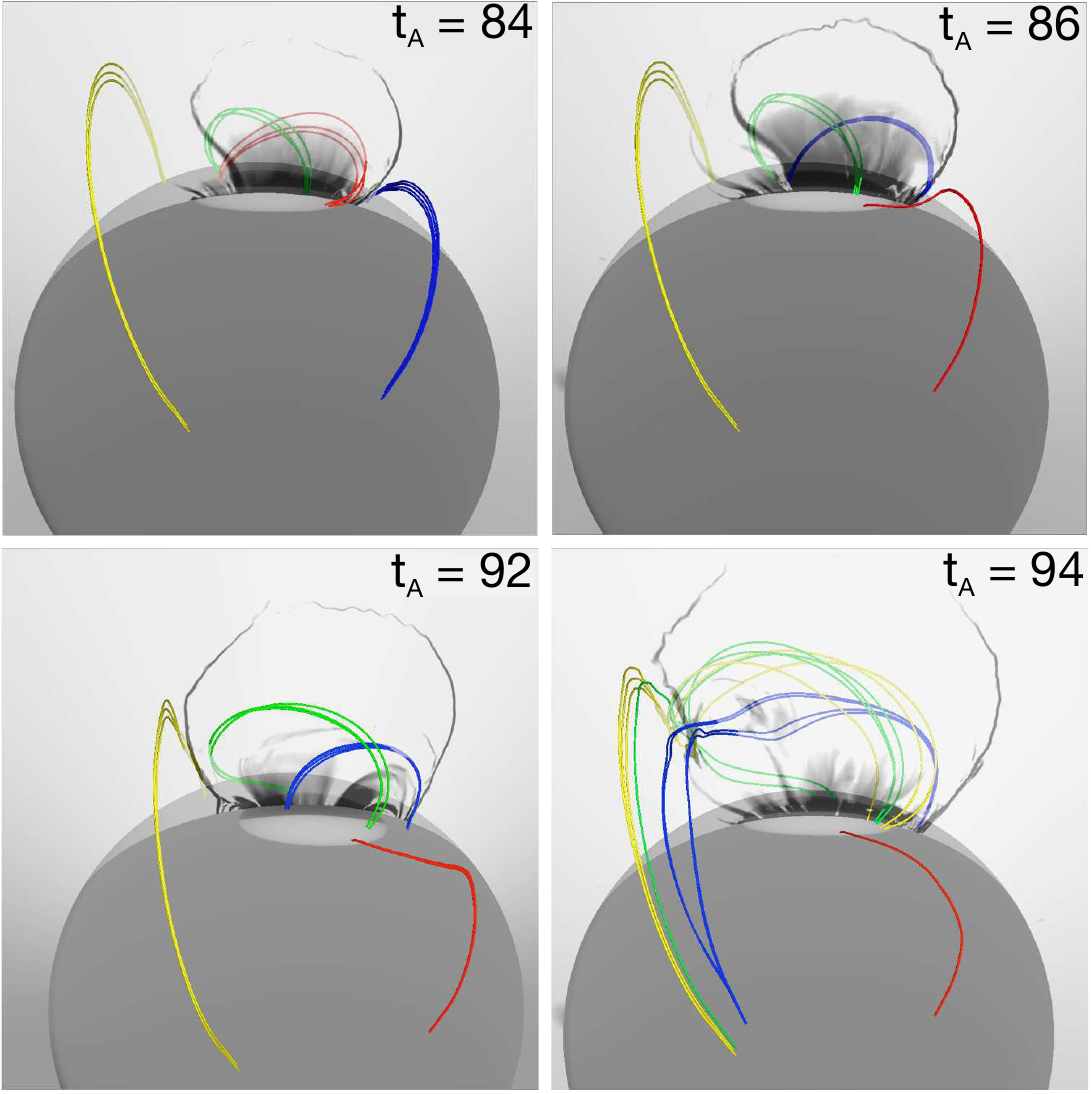}
 }
\caption{Reconnection on the sides of the fan surface. The radial magnetic field is color-coded in greyscale on the solar surface. Coloured field lines show the connectivity evolution of the magnetic field. Blue field lines are plotted from fixed footpoints to the west of the active region; yellow field lines are anchored in the southern hemisphere close to the pole. Green and red field lines, respectively, are plotted from footpoints advected by the photospheric flows in the negative north and positive south polarity of the active region. In each panel, the current density in the $(r,\phi)$ plane is greyscale-coded and shown in transparency.}
  \label{fig10}
\end{figure}

\begin{figure}
\centerline{
 \includegraphics[width=0.6\textwidth,clip=]{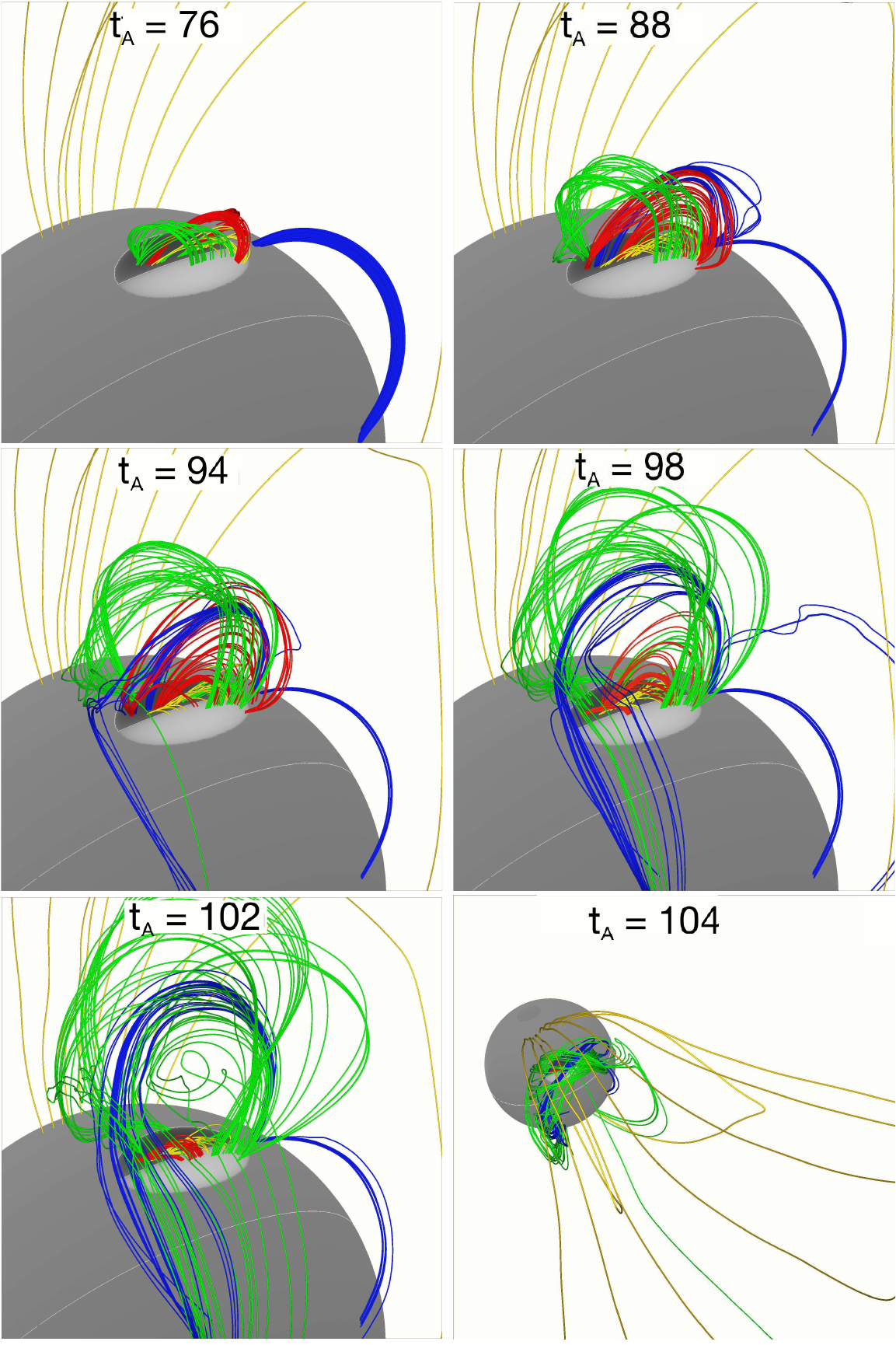}
 }
\caption{3D evolution of the magnetic field of the CME. Panels a) to e) show a zoomed view of the connectivity evolution of the magnetic field lines during the CME initiation and eruption. Field lines are coloured to represent the different connectivity domains. Yellow field lines are anchored near the north pole and initially open into the interplanetary medium. Green field lines are initially rooted from footpoints in the south positive polarity in the area advected by the photospheric flow; red and yellow field lines are rooted in the photospheric flow in the negative north polarity. Blue field lines are initially rooted on the west side of the active region, outside of the photospheric flow. Panel f) shows a large-scale view of the CME eruption. An animation of this figure is provided online.}
  \label{fig11}
\end{figure}

\begin{figure}
\centerline{
 \includegraphics[width=0.9\textwidth,clip=]{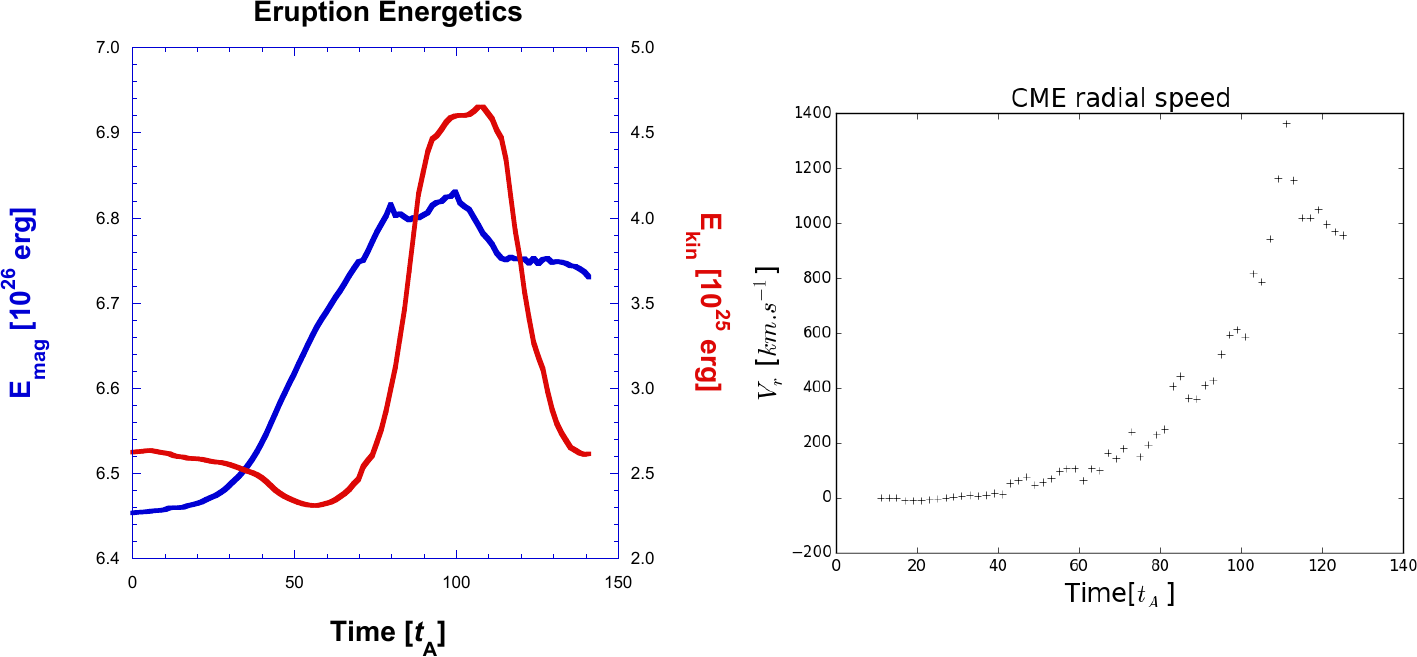}
 }
 \caption{Energetics of the CME. Left panel shows the evolution of the
   magnetic energy (blue curve) and the kinetic energy (red curve)
   from the onset of the photospheric driving ($ t_A= 0 $) to the end
   of the simulation ($ t_A= 141 $). Right panel displays the
   evolution of the CME radial velocity computed at the apex of the
   closed loops below the null point. The data points for the CME stop
   after $ t_A= 121 $, when the flux rope begins to reconnect with its
   environment.}
  \label{fig12}
\end{figure}

\begin{figure}
\centerline{
 \includegraphics[width=0.6\textwidth,clip=]{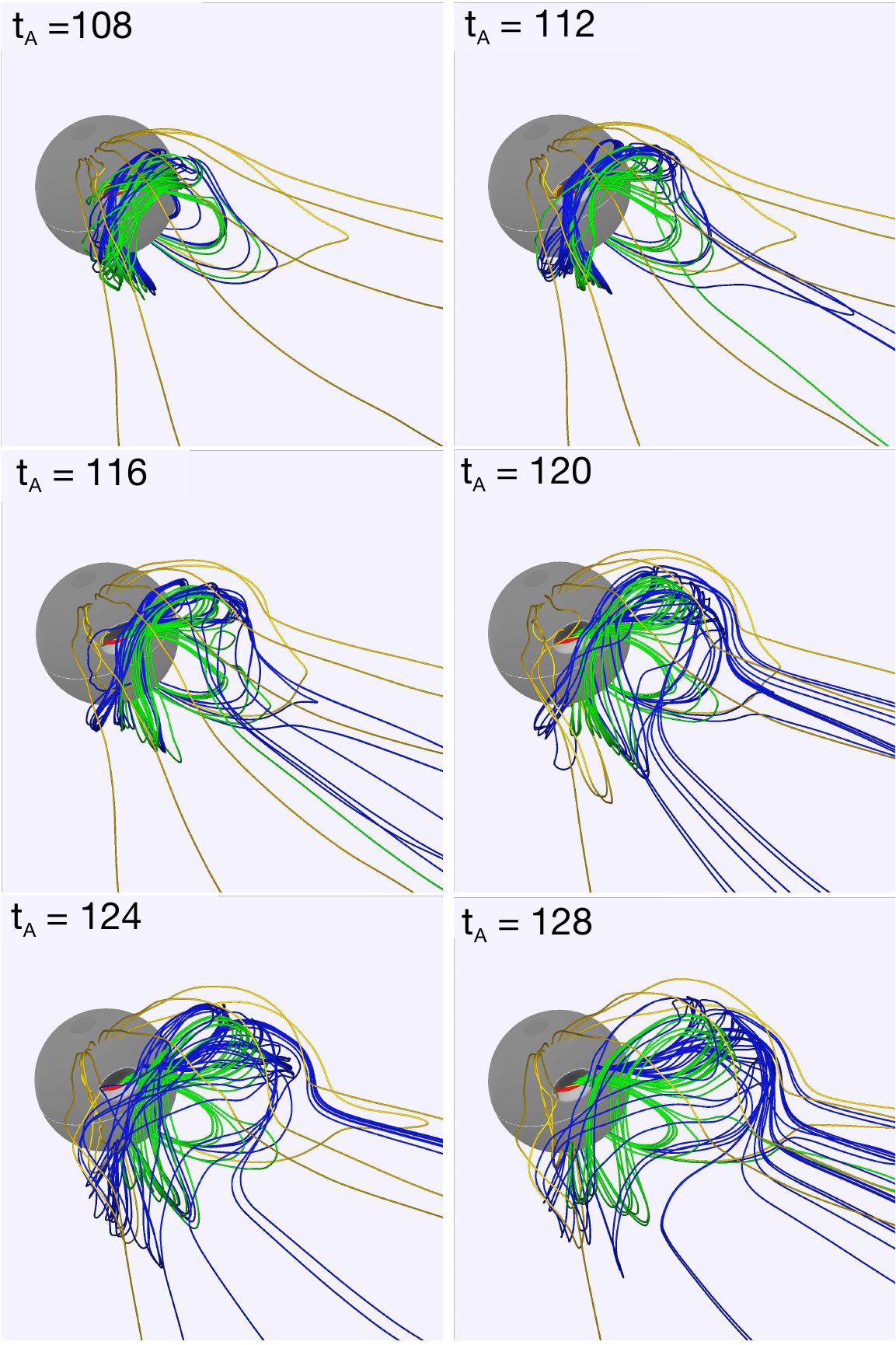}
 }
\caption{Interaction of the CME with the interplanetary magnetic field. Field lines are plotted with the same color code as Figure~\ref{fig11}. Dark blue field lines show the magnetic flux that reconnects with the open magnetic field of the interplanetary medium. As the time advances (from left to right and top to bottom), more dark blue field lines open into the interplanetary medium. An animation of this figure is provided online.}
  \label{fig13}
\end{figure}

\newpage

\end{document}